\documentclass[iop,revtex4]{emulateapj}
\usepackage{amstext}
\usepackage[breaklinks,colorlinks,citecolor=blue,backref=section]{hyperref}
\usepackage[all]{hypcap}

\usepackage[english]{babel} 
\usepackage[utf8]{inputenc} 
\usepackage[T1]{fontenc}
\usepackage[mathscr]{eucal}
\usepackage{graphicx}
\usepackage{amsmath}
\usepackage{epstopdf}
\usepackage{epsf,color}
\usepackage{natbib}
\newcommand\blfootnote[1]{%
  \begingroup
  \renewcommand\thefootnote{}\footnote{#1}%
  \addtocounter{footnote}{-1}%
  \endgroup
}

\usepackage{multirow}
\usepackage{longtable,lscape}
\usepackage{txfonts}
\usepackage{amsfonts}
\usepackage{amssymb}
\usepackage{newlfont}
\usepackage{textcomp}
\usepackage{times}
\usepackage{units}
\usepackage{mathrsfs}
\shorttitle{Dissecting the high-z interstellar Medium through intensity mapping cross-correlations}
\shortauthors{Serra P. {\it et al}.}
\begin{document}
\title{Dissecting the high-z interstellar Medium through intensity mapping cross-correlations}
\author{Paolo Serra\altaffilmark{1,2}, Olivier Dor\'e\altaffilmark{1,2}, Guilaine Lagache\altaffilmark{3}}
\email{Paolo.Serra@jpl.nasa.gov}
\altaffiltext{1}{Jet Propulsion Laboratory, California Institute of Technology, Pasadena, California 91109, USA, \email{Paolo.Serra@jpl.nasa.gov}}
\altaffiltext{2}{California Institute of Technology, Pasadena, California, 91125 USA}
\altaffiltext{3}{Aix Marseille Universit\'e, CNRS, LAM (Laboratoire d'Astrophysique de Marseille) UMR 7326, 13388, Marseille, France}
%\date{}

\begin{abstract}
We explore the detection, with upcoming 
spectroscopic surveys, of three-dimensional power 
spectra of emission line fluctuations produced in different phases of 
the Interstellar Medium (ISM) by forbidden 
transitions of ionized carbon [CII] (157.7 $\mathrm{\mu}$m), 
ionized nitrogen [NII] (121.9 $\mathrm{\mu}$m and 205.2 $\mathrm{\mu}$m)
and neutral oxygen [OI] (145.5 $\mathrm{\mu}$m) at redshift $\mathrm{z>4}$. 
%The emission line [CII] from ionized carbon at 
%157.7 $\mathrm{\mu}$m, and multiple emission lines 
%from carbon monoxide, are the main targets of 
%recently planned ground-based surveys, and an important 
%foreground for future space-based surveys like the
%Primordial Inflation Explorer (PIXIE). However, other 
%far-infrared lines, such as the oxygen [OI] 
%(145.5 $\mathrm{\mu}$m) line, and the nitrogen 
%[NII] (121.9 $\mathrm{\mu}$m and 205.2 $\mathrm{\mu}$m) lines, 
%%might be detected in correlation with [CII] with 
%reasonable signal-to-noise ratio (SNR).
These lines are important coolants of both the neutral 
and the ionized medium, and probe multiple
phases of the ISM. In the framework of the halo model, 
we compute predictions of the three-dimensional 
power spectra for two different surveys, 
showing that they have the required 
sensitivity to detect cross-power spectra between the [CII]
line and both the [OI] line and the [NII] lines with sufficient SNR.
The importance of cross-correlating multiple lines with the 
intensity mapping technique is twofold. 
On the one hand, we will have multiple probes of the different 
phases of the ISM, which is key to understand the 
interplay between energetic sources, 
and the gas and dust at high redshift. 
This kind of studies will be useful 
for a next-generation space observatory 
such as the NASA Far-IR Surveyor, that will probe 
the global star formation and the ISM of galaxies 
from the peak of star formation to the epoch of reionization. 
On the other end, 
emission lines from 
external galaxies are an important foreground when 
measuring spectral distortions
of the Cosmic Microwave Background (CMB) spectrum with 
future space-based experiments like PIXIE; 
measuring fluctuations in the intensity mapping regime will help 
constraining the mean 
amplitude of these lines, and will allow us to better handle this 
important foreground.
\end{abstract}

\maketitle

%===============================================================
\section{Introduction}
\label{sect:intro}
\blfootnote{Copyright 2016. All rights reserved.}
\noindent Intensity mapping, introduced in \cite{madau1997,suginohara1999,shaver1999}, is an observational 
technique for measuring brightness 
fluctuations of emission lines 
produced by sources below the detection limit. 
Atomic and molecular emission lines, produced 
at a given redshift, are observed as 
fluctuations redshifted at a certain 
frequency, enabling us to 
map the three-dimensional structure of the 
Universe and compute, 
for each redshift slice, statistical quantities of 
interest such as the power spectrum. Intensity mapping, by measuring 
the aggregate radiation emitted by all galaxies in a given 
redshift slice, does not suffer from the incompleteness problem, 
while traditional galaxy surveys, being flux-limited, do not detect the faintest galaxies. This can 
be a serious disadvantage 
if the galaxy luminosity function has a sufficiently steep end, 
as shown in \cite{uzgil2014}. \\
One of the first and main targets of intensity mapping is the 21 cm neutral 
hydrogen line \citep{battye2004,chang2010,bull2015}, which, in principle, opens a new window 
on both the formation of structures at high redshift 
and the history of reionization 
\citep{furlanetto2006}. However, lines from other atoms and molecules 
can be used to constrain the physics of the ISM 
in a broad redshift range.\\ 
The carbon [CII] fine-structure line at 157.7 $\mathrm{\mu}$m, 
arising from the $\mathrm{^2P_{3/2}\rightarrow\,^2P_{1/2}}$ 
fine-structure transition, is one 
of the most promising lines not only to understand 
star-formation in galaxies  
\citep{boselli2002,delooze2011,delooze2014,herrera2015}, 
but also to constrain the epoch of reionization and the physics of the ISM 
\citep{gong2011a,gong2011b,uzgil2014,silva2015,lidz2016,cheng2016}.\\
Both theory and observations indicate that 
the atomic [CII] fine-structure is the dominant 
coolant of the neutral ISM 
\citep{hollenbach1999,bsalas2012}, and one 
of the brightest lines in the Spectral
Energy Distribution (SED) of a typical star-forming galaxy, 
with luminosities ranging from $0.01\%$ to 
$1\%$ of the total infrared luminosity 
\citep{stacey1991,maiolino2005,iono2006,maiolino2009,stacey2010,ivison2010,wagg2010,
debreuck2011}. In fact, 
carbon is the forth most abundant element in the Universe. It has 
a low ionization potential, only 11.26~eV
(see Table~\ref{tab:spinoglio}), below the 13.6 eV of hydrogen 
ionization; this  ensures it is present both in the ionized
and in the neutral medium. Moreover, the [CII] fine-structure transition of 
ionized carbon is characterized by a low temperature (91~K), and low 
critical density for collisions with hydrogen\footnote{The critical density for an excited state is the density for 
which collisional deexcitation equals radiative deexcitation, 
see \cite{draine2011}.}.\\
Intensity mapping from the rotational transitions of carbon monoxide, 
and, in particular, the lowest order transition CO(1-0) at 115 GHz, 
have also received increased attention in the past 
few years. Carbon monoxide emission lines at a given 
redshift act as a foreground contamination both 
for CMB observations \citep{righi2008,dezotti2016}, 
and for [CII] intensity mapping 
surveys targeting background galaxies at 
higher redshifts \citep{gong2011b,lidz2016,cheng2016}. 
Carbon monoxide molecules are easily produced 
from carbon and oxygen in 
star-forming regions, and CO intensity mapping 
provides information on 
the spatial distribution and redshift evolution 
of star formation in the Universe \citep{visbal2010,carilli2011,lidz2011,gong2011a,pullen2013,
breysse2014}.\\
At far-infrared (FIR) frequencies, many other lines can in principle be 
targeted by intensity mapping surveys, such as [OI] 
(63 $\mathrm{\mu}$m and 145 $\mathrm{\mu}$m), 
[NII] (122 $\mathrm{\mu}$m and 205 $\mathrm{\mu}$m), [OIII] 
(52 $\mathrm{\mu}$m and 88 $\mathrm{\mu}$m), and 
[CI] (610 $\mathrm{\mu}$m and 371 $\mathrm{\mu}$m), while   
proposed lines in other frequency bands include measurements of HeII 
(0.164 $\mathrm{\mu}$m) to constrain properties of Population III stars 
\citep{visbal2015}, Ly$\mathrm{\alpha}$ (0.1216 $\mathrm{\mu}$m) to probe 
reionization and star formation \citep{pullen2014}, 
and OII (0.3737 $\mathrm{\mu}$m) and H$\mathrm{\alpha}$ (0.6563 $\mathrm{\mu}$m)
to study the large scale clustering at redshifts $\mathrm{1<z<4}$ 
\citep{fonseca2016}.\\
As emphasized in \cite{lidz2016}, the sensitivity of intensity mapping 
measurements will rapidly increase in the near future, thanks to 
advances in detector technology, and some surveys
are already in progress, or have been planned,
to perform intensity mapping of one or more emission lines from
sources at multiple redshifts. The CO Power Spectrum Survey (COPPS)
\citep{keating2015} recently published measurement of the CO
abundance and power spectrum from the CO(1-0) transition in the redshift
range $\mathrm{2.3<z<3.3}$ \citep{keating2016}, and the Carbon Monoxide
Mapping Array Pathfinder (COMAP see \cite{li2016})
has been proposed to study the CO emission at similar redshifts.\\
Experiments targeting the [CII] emission line include the 
Tomographic Ionized-Carbon Mapping Experiment 
(TIME-Pilot, \cite{crites2014}), and CONCERTO (CarbON CII line 
in post-rEionization and ReionizaTiOn epoch, Lagache {\it et al}., 
in preparation), 
while the Spectrophotometer for the History of the Universe, Epoch of
Reionization, and Ice Explorer (SPHEREx) will focus on
Ly$\mathrm{\alpha}$, Ly$\mathrm{\beta}$ and [OIII] \citep{dore2014,dore2016}. 
The Cryogenic-Aperture Large 
Infrared-Submillimeter Telescope Observatory (CALISTO) 
\citep{bradford2015} has been proposed to measure, among other things, 
multiple FIR fine-structure transitions such as [NeII], [OI], 
[OIII] and, for $z<2$, [CII].\\
Foregrounds are an important concern for intensity mapping surveys. 
Apart from the continuum emission from our Galaxy, a survey
targeting an emission line observed 
at a given frequency $\nu_{\mathrm{obs}}$ 
will also detect the sum of emissions of N atoms or 
molecules $\mathrm{\alpha_i}$ coming from redshifts $z_i$, 
whose lines are redshifted to the same observed frequency, 
so that the measured intensity $\mathrm{I^{\nu_{obs}}}$ 
can be written as:
\begin{eqnarray}
I^{\nu_{\mathrm{obs}}} &=& \sum_{i=1}^NI_{i}^{\nu^i_{\mathrm{em}}/(1+z_i)}(\alpha_i, z_i).
\end{eqnarray}   
Different methods to overcome this difficulty have been 
proposed so far. Some authors 
\citep{visbal2011,breysse2015,silva2015} explore the possibility 
of mitigating this contamination by progressively masking 
the brightest pixels in the observed map. However, when dealing with 
[CII] maps at very high redshift (e.g. $\mathrm{z=7}$), 
a percentage of the signal will be masked in the process, 
and such a loss of information 
translates in a underestimation of the amplitude of the measured power 
spectrum \citep{breysse2015}. This is unfortunate because, 
while the cosmological information content of the 
measured power spectrum is mainly encoded in its {\it shape} (primordial 
non-Gaussianity, neutrino masses, modified gravity can all be tested by 
looking at the shape of the clustering power spectrum), most of the meaningful 
astrophysical processes are constrained by the {\it amplitude} of the 
spectrum.
Another method, recently discussed 
in \cite{lidz2016,cheng2016}, exploits the fact that the interloper lines, 
being emitted at different redshifts respect to the targeted line, 
will introduce an anisotropic component in the power spectra, due to the 
incorrect redshift projection.\\
A third method to mitigate contamination from different lines 
has been proposed by \cite{visbal2010,visbal2011}, and involves the 
cross-correlation between maps measured at different 
frequencies, whose emission comes from atoms and molecules at the same 
redshift. Since all contaminant lines in each map will generally come from 
different redshifts, they will not contribute to the signal in the 
cross-correlation, but they will only add noise to the measurement. 
While cross-correlation measurements are generally more complicated to be 
carried out, most surveys proposed so far work in a broad frequency range, 
and multiple cross-correlations produced at the same redshift among 
lines from different atoms and 
molecules might be attempted, at least in the non-linear  
regime. If the amplitudes of the lines to be 
cross-correlated is large enough, the information content 
from these measurements will be vast, and it will enable us 
to constrain various physical processes of the ISM.\\
In this paper we propose the use of cross-correlation 
measurements among various emission lines from carbon, 
oxygen, and nitrogen to constrain the mean amplitude 
of each emission line at redshift $z>4$. 
Using measurements of the Cosmic Infrared Background (CIB) 
angular power spectra from \textit{Herschel}/SPIRE \citep{viero2013a} and 
{\it Planck} \citep{planckXXX}, 
coupled to a compilation of star formation rate density (SFRD) measurements from 
\cite{madau2014}, we constrain the galaxy FIR luminosity as a 
function of the halo mass at all relevant redshifts. 
By using scaling relations from \cite{spinoglio2012} 
to link the intensity of emission lines to the 
constrained galaxy infrared luminosity, we compute 3D emission 
line power spectra for all relevant lines. Focusing on two 
experimental setups, corresponding to present and 
future ground-based surveys, we show that 
multiple cross-correlations with the [CII] line can constrain 
the mean amplitudes of all lines. This is important not only 
to constrain average properties of the ISM of galaxies at high redshift, 
but also because, as shown in
\cite{mashian2016,dezotti2016,carilli2016}, especially the CO and [CII] line 
emission from galaxies across cosmic time distort the 
CMB spectrum at a level that must be taken into account 
by future space-based surveys aiming at measuring the tiny 
spectral distortions of the CMB, such as PIXIE. Intensity 
mapping, by constraining the mean amplitude of the signal, 
will healp dealing with 
this important foreground.\\
In Sect.~\ref{sect:halo} we will derive the formalism used to compute 
emission line power spectra from the Halo model. 
We will then discuss in Sect.~\ref{sect:ism} the physics of the ISM in the context of emission 
lines from carbon, oxygen, and nitrogen, with particular focus on 
all possible cross-correlations to be performed using the experimental 
setups discussed in Sect.~\ref{sect:exp}. Finally we will discuss our 
main results in Sect.~\ref{sect:discuss}.\\
Throughout this paper, we adopt the standard flat 
$\mathrm{\Lambda}$CDM model as our fiducial background cosmology, with 
parameter values derived from the best-fit model of the CMB power spectrum 
as measured by \cite{planck2014}. 

\section{A Halo model for emission line amplitudes}
\label{sect:halo}
The computation of 3D auto- and cross-power spectra of intensity line 
emission is performed in the context of a Halo model developed in 
\cite{shang2012}, where the galaxy luminosity is linked to the 
mass of the host dark matter halo with a simple parameteric form. 
It has been successfully 
applied to the interpretation of the latest measurements of 
angular CIB power spectra from {\it Herschel}/SPIRE \citep{viero2013a} and 
{\it Planck} \citep{planckXXX}.\\
Using the latest measurements of CIB auto- and cross-power 
spectra at 250, 350 and 500 $\mathrm{\mu}$m from \cite{viero2013a}, 
together with a compilation of measurements
of SFRD in the redshift
range $\mathrm{0<z<6}$ \citep{madau2014}, we are able to constrain the 
galaxy infrared luminosity as a function of halo mass and redshift.
We then use known scaling relations from \cite{spinoglio2012} 
to compute the amplitudes of emission lines from carbon, oxygen, 
and nitrogen with respect to the constrained galaxy infrared luminosity. This 
allows us to compute the amplitudes of 3D power spectra for 
all relevant emission lines at all redshifts. This approach is very similar 
to that discussed in \cite{cheng2016}. \\
%In the following we will briefly introduce the 
%notation used and the analysis performed to constrain 
%the galaxy infrared luminosity as a function 
%of mass and redshift, in the context of a halo model that will be subsequently used to make 
%predictions for the 3D power spectra of emission lines.
\subsection{The Halo model for CIB anisotropies}
The halo model is a phenomenological description of the galaxy
clustering at all angular scales \citep{cooray2002}.
Assuming that all galaxies live in virialized dark matter structures,
called halos, and using a recipe to populate halos
with galaxies, the clustering power spectrum results from the sum
of two components: a 1-halo term, related to correlations between 
galaxies in the same halo, and responsible for the clustering 
at small angular scales, and a 2-halo term,
which describes the power spectrum at large angular scales,
and is due to correlations between galaxies belonging to separated
dark matter halos. \\
The angular power spectrum of CIB anisotropies, observed at frequencies 
$\mathrm{\nu}$ and $\mathrm{\nu^{\prime}}$, is defined as:
\begin{eqnarray}
\label{eqn:ang1}
\langle\delta\,I_{lm,\nu}\delta\,I_{l^{\prime}m^{\prime},\nu^{\prime}}\rangle &=& 
C_{l,\nu\nu^{\prime}}
\delta_{\nu\nu^{\prime}}\delta_{mm^{\prime}}
\end{eqnarray}
where $\mathrm{I_{\nu}}$ 
is the specific intensity at that frequency, given by:
\begin{eqnarray}
\label{eqn:ang2}
I_{\nu}(z) &=& \int\,dz\frac{d\chi}{dz}aj(\nu,z) \\
\nonumber
&=& \int\,dz\frac{d\chi}{dz}\bar{j}(\nu,z)
\Big(1+\frac{\delta\,j(\nu,z)}{\bar{j}(\nu,z)}\Big);
\end{eqnarray}
here $\mathrm{\chi(z)}$ denotes the comoving distance at redshift z, 
$\mathrm{a(z)}$ is the scale factor, and $\mathrm{j(\nu,z)}$ 
is the comoving emission coefficient.\\ 
In Limber approximation \citep{limber1954}, Eqs.~\ref{eqn:ang1} and 
\ref{eqn:ang2} can be combined to give the clustering angular 
power spectrum as:
\begin{eqnarray}
\label{eqn:clustering}
C^{\nu\nu^{\prime}}_{\mathrm{clust}}(l) &=& \int\frac{dz}{\chi^2}\frac{d\chi}{dz}a^2(z)\bar{j}(\nu,z)
\bar{j}(\nu^{\prime},z)P^{\nu\nu^{\prime}}(k=l/\chi,z), 
\end{eqnarray}
where $\mathrm{P^{\nu\nu^{\prime}}(k,z)}$ is the 3D 
power spectrum of the emission coefficient, expressed as:
\begin{eqnarray}
\label{eqn:3dpower}
\langle\delta\,j(\vec{k},\nu)\delta\,
j(\vec{k}^{\prime},\nu^{\prime})\rangle &=&
(2\pi)^3\bar{j}_{\nu}\bar{j}_{\nu^{\prime}}
P^{\nu\nu^{\prime}}_j\delta^3(\vec{k}-\vec{k^{\prime}}).
\end{eqnarray}
This term is composed by the mentioned 1-halo and 2-halo components. 
Thus, together with a 
scale independent shot-noise power spectrum, describing the contribution 
from random fluctuations due to the Poisson distribution of sources, 
the total CIB angular power spectrum is:
\begin{eqnarray}
\label{eqn:ang_power_total}
C^{\nu\nu^{\prime}}_{\mathrm{tot}}(l) &=& C^{\nu\nu^{\prime}}_{\mathrm{1h}}(l) + C^{\nu\nu^{\prime}}_{\mathrm{2h}}(l) + C^{\nu\nu^{\prime}}_{\mathrm{SN}}(l).
\end{eqnarray}
This quantity will be computed and fit to {\it Herschel}/SPIRE measurements 
of CIB angular power spectra in order to constrain the galaxy infrared 
luminosity.\\
Below we show how to compute the two clustering terms. 
This formalism will be useful in Sect.~\ref{sect:3dImapping}, 
when computing 3D power spectra of emission lines.\\
The mean emissivity $\mathrm{\bar{j}_{\nu}(z)}$ from all galaxies is computed from the infrared 
galaxy luminosity function $\mathrm{dn/dL}$ as:
\begin{eqnarray}
\label{eqn:j0}
\bar{j}_{\nu}(z) &=& \int\,dL \frac{dn}{dL}(L,z)
\frac{L_{(1+z)\nu}(M,z)}{4\pi},
\label{eq_jnu}
\end{eqnarray}
where the galaxy luminosity $\mathrm{L_{(1+z)\nu}}$ is observed at the 
frequency $\nu$ with a flux given by:
\begin{eqnarray}
\mathrm{S_{\nu}} &=& \mathrm{\frac{L_{\nu(1+z)}}{4\pi\chi^2(z)(1+z)}}.
\end{eqnarray}
Neglecting any scatter between galaxy luminosity and dark matter halo mass,
the luminosity of central and satellite galaxies can be expressed as
$\mathrm{L_{{\mathrm{cen}},(1+z)\nu}(M_{\mathrm{H}},z)}$ and
$\mathrm{L_{{\mathrm{sat}},(1+z)\nu}(m_{\mathrm{SH}},z)}$,
where $\mathrm{M}_{\mathrm{H}}$ and $\mathrm{m}_{\mathrm{SH}}$ denote the halo
and sub-halo masses, respectively. We can thus rewrite
Eq.~\ref{eq_jnu} as the sum of the contributions 
from central and satellite galaxies as:
\begin{eqnarray}
\label{eqn:j1}
\bar{j}_{\nu}(z)&=& \int dM \frac{dN}{dM}(z)\frac{1}{4\pi}
 \Big\{\frac{}{}N_{\mathrm{cen}}L_{{\mathrm{cen}},(1+z)\nu}(M_{\mathrm{H}},z)\\
\nonumber
& &+\int dm_{\mathrm{SH}} \frac{dn}{dm}(m_{\mathrm{SH}},z)
 L_{{\mathrm{sat}},(1+z)\nu}(m_{\mathrm{SH}},z)\Big\};
\end{eqnarray}
here $\mathrm{dN/dm}$ \citep{tinker2008} and $\mathrm{dn/dm}$ 
\citep{tinker2010} denote the halo and sub-halo mass function respectively, 
while $N_{\mathrm{cen}}$ is the number of central galaxies in a halo, 
which will be assumed equal to zero if the mass of the host halo is lower
than $M_{\mathrm{min}} = 10^{11}$M$_{\odot}$, and one otherwise.\\
Introducing $f^{\mathrm{cen}}_{\nu}$ and $f_{\nu}^{\mathrm{sat}}$ as the 
number of central and satellite galaxies weighted by their luminosity, as:
\begin{eqnarray}
f_{\nu}^{\mathrm{cen}}(M,z) = N_{\mathrm{cen}}
 \frac{L_{{\mathrm{cen}},(1+z)\nu}(M_{\mathrm{H}},z)}{4\pi},
\label{eqn:fcen}
\end{eqnarray}
and
\begin{eqnarray}
\label{eqn:fsat}
f_{\nu}^{\mathrm{sat}}(M,z) &=& \int_{M_{\mathrm{min}}}^{M}dm
 \frac{dn_{}}{dm}(m_{\mathrm{SH}},z|M) \\\nonumber
&& \times \frac{L_{{\mathrm{sat}},(1+z)\nu}(m_{\mathrm{SH}},z)}{4\pi},
\end{eqnarray}
the power spectrum coefficient of CIB anisotropies at the observed frequencies
$\nu$ and $\nu^\prime$ can be written as the sum of a 1-halo term and 
2-halo term as, respectively:
\begin{eqnarray}
\label{eqn:pj1h}
P^{}_{{\mathrm{1h}},\nu\nu^{\prime}}(k,z)&=&
 \frac{1}{\bar{j}_{\nu}\bar{j}_{\nu^\prime}}\int_{M_{\mathrm{min}}}^{\infty}dM
 \frac{dN}{dM}\\\nonumber
&&\times\, \left\{f_{\nu}^{\mathrm{cen}}(M,z)f_{\nu^\prime}^{\mathrm{sat}}(M,z)u(k,M,z)
\right. \\\nonumber
&&\quad +f_{\nu^{\prime}}^{\mathrm{cen}}(M,z)f_{\nu}^{\mathrm{sat}}(M,z)
 u(k,M,z)\\\nonumber
&&\left.\quad +f_{\nu}^{\mathrm{sat}}(M,z)f_{\nu^{\prime}}^{\mathrm{sat}}(M,z)
 u^2(k,M,z)\right\},\\
\label{eqn:pj2h}
P^{}_{{\mathrm{2h}},\nu\nu^{\prime}}(k,z)&=&
 \frac{1}{\bar{j}_{\nu}\bar{j}_{\nu^\prime}}D_{\nu}(k,z)
 D_{\nu^{\prime}}(k,z)P_{\mathrm{lin}}(k,z),
\end{eqnarray}
where
\begin{eqnarray}
D_{\nu}(k,z)&=&\int_{M_{\mathrm{min}}}^{\infty}dM\frac{dN}{dM}b(M,z)u(k,M,z)\\
\nonumber
&&\times\, \left\{f_{\nu}^{\mathrm{cen}}(M,z)+f_{\nu}^{\mathrm{sat}}(M,z)\right\},
\label{eqn:pjf}
\end{eqnarray}
and $\mathrm{u(k,M,z)}$ is the Fourier transform of the 
Navarro-Frenk-White (NFW) density profile \citep{navarro1997}, 
with concentration parameter from \cite{duffy2010}. 
The term $\mathrm{b(M,z)}$ denotes the halo bias
\citep{tinker2010}. The linear dark matter
power spectrum $P_{\mathrm{lin}}(k)$ is computed using 
{\tt CAMB} (\url{http://camb.info/}). \\
The final ingredient to be specified is the link between galaxy
luminosity and host dark matter halo mass. Following \cite{shang2012},
we assume a parametric function, where the dependence of the
galaxy luminosity on frequency, redshift, and halo mass is factorized in
three terms as:
\begin{eqnarray}
\label{eqn:luminosity}
L_{(1+z)\nu}(M,z)= L_0 \Phi(z) \Sigma(M) \Theta[(1+z)\nu].
\label{eqn:lfunc}
\end{eqnarray}
The parameter $\mathrm{L_0}$ is a free normalization parameter 
whose value is set by the amplitude of both the CIB power 
spectra and the SFRD. It has no physical 
meaning, and it will not be discussed further in
the rest of the paper.\\
%We further define $\mathrm{Q(M,z)}$ as:
%\begin{equation}
%Q(M,z) \equiv L_0 \Phi(z) \Sigma(M,z);
%\end{equation}
%this quantity will be useful in the rest of the paper.\\
A very simple functional form 
\citep[see][and reference therein]{blain2002} is assumed 
for the galaxy SED:
\begin{eqnarray}
\Theta (\nu) \propto
\left\{\begin{array}{ccc}
\nu^{\beta}B_{\nu}\,(T_{\mathrm d})&
 \nu<\nu_0\, ;\\
\nu^{-2}&  \nu\ge \nu_0\,,
\end{array}\right.
\label{eqn:thetanu}
\end{eqnarray}
where $\mathrm{T_d}$ is the dust temperature 
averaged over the redshift range considered, and 
$\mathrm{\beta}$ is the emissivity of the 
Planck function $\mathrm{B_{\nu}(T_d)}$. We note that 
we discarded a redshift dependence of the dust temperature, because it is 
not very well constrained by the data. The power-law function at frequencies 
$\mathrm{\nu\geq\nu_0}$ has been found more in
agreement with observations than the exponential Wien tail 
(see also \cite{hall2010,viero2013a,shang2012,planckXXX}).
We also assume a redshift-dependent, global normalization of the L--M
relation of the form
\begin{eqnarray}
\Phi (z)= \left(1+z\right)^{\delta}.
\label{eqn:phiz}
\end{eqnarray}
As explained in \cite{shang2012}, a power law is motivated by the 
study of the star formation rate (SFR) per unit 
stellar mass, or specific star formation rate (sSFR). Assuming 
that the stellar mass to halo mass ratio does not 
evolve substantially with redshift, the ratio of 
galaxy infrared luminosity $\mathrm{L_{IR}}$ to halo mass has an evolution 
similar to the sSFR, thanks to the correlation 
between SFR and infrared luminosity \citep{kennicutt1998}.\\
Finally, following \cite{shang2012,viero2013a,planckXXX} we assume a 
log-normal function for the L-M relation, as:
\begin{eqnarray}
\Sigma(M) = M \frac{1}{(2\pi\sigma^2_{L/M})^{0.5}}\mathrm{exp}\Big[-\frac{(\mathrm{log}_{10}M - \mathrm{log}_{10}M_{\mathrm{eff}})^2}{2\sigma_{L/M}^2}\Big],
\end{eqnarray}
where $\mathrm{M_{eff}}$ describes the most efficient halo 
mass at hosting star formation, while 
$\sigma_{L/m}$ accounts for the range of halo 
masses mostly contributing to the infrared luminosity. 
Such a functional form captures the fact that, for halo masses much 
lower and much higher than $\mathrm{M_{eff}}$, 
various mechanisms prevent an efficient star formation 
\citep{benson2003,silk2003,bertone2005,croton2006,dekel2006,bethermin2012,behroozi2013}.

\subsection{Analysis}
\begin{figure*}[!t]
\begin{center}
%\begin{array}{ccc}
\includegraphics[width=150mm]{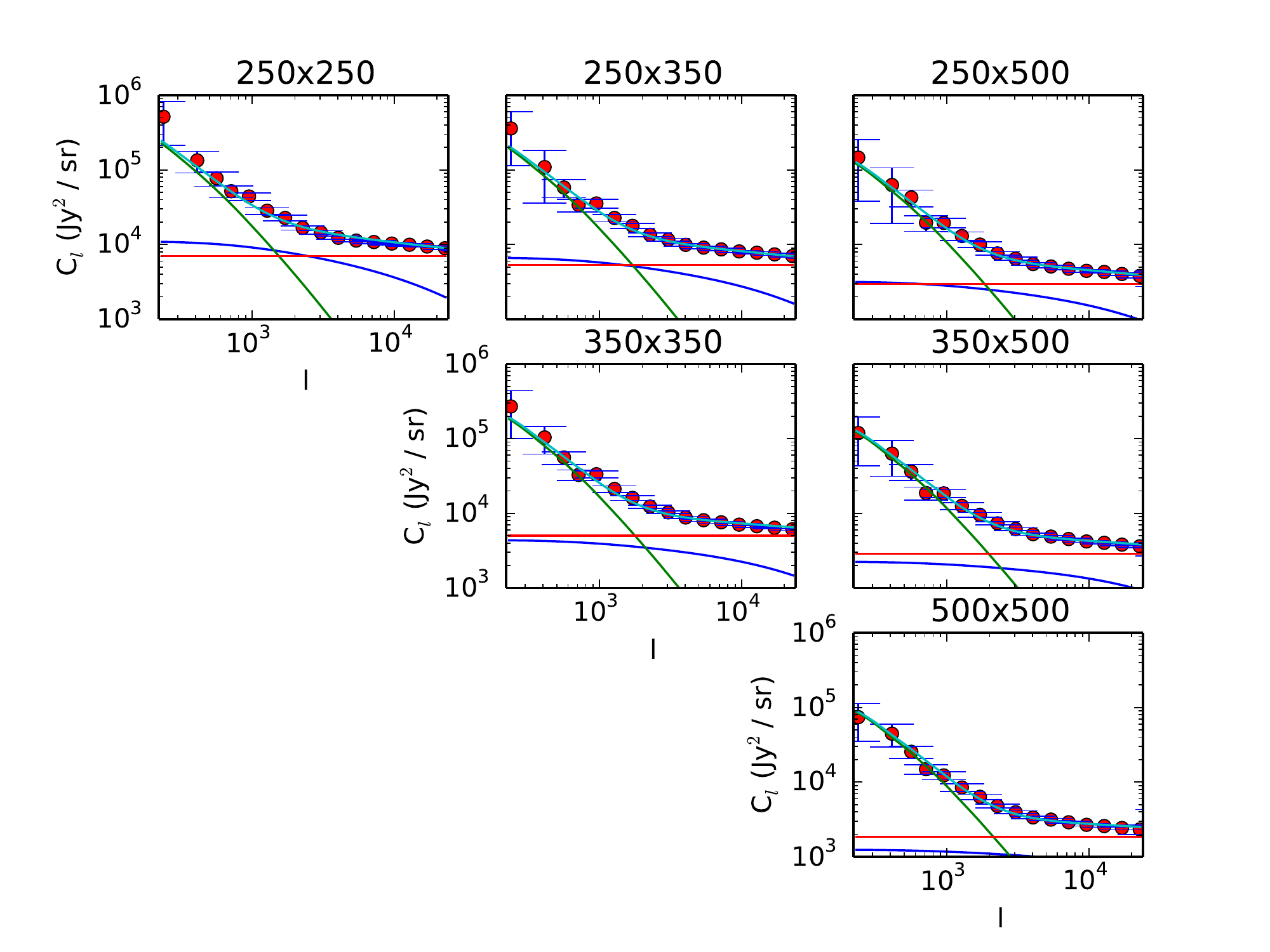}
\end{center}
\caption{Angular CIB auto- and cross-power spectra at 250, 350, 500 
$\mathrm{\mu}$m from \textit{Herschel}/SPIRE, together with the best-fit curves for the 
1-halo (Blue line), 2-halo (Green line), shot-noise (Red line) and total power spectra 
(Cyan line).} 
\label{fig:power_spectrum}
\end{figure*}

We perform a Monte Carlo Markov Chain (MCMC) analysis of the
parameter space, using a modification of the publicly available
code {\tt CosmoMC} \citep{lewis2002}, and fitting to six 
CIB auto- and cross-power spectra from \cite{viero2013a} in the multipole 
range $\mathrm{200<l<23000}$. We also add 
a dataset for the SFRD as a function of 
redshift by averaging multiple measurements, discussed in 
\cite{madau2014}, in eleven redshift bins in the range $\mathrm{0<z<6}$.\\
We vary the following set of parameters:
\begin{eqnarray}
\mathscr{P} \equiv \{M_{\mathrm{eff}}, T_d, \delta, L_\mathrm{0} \},
\end{eqnarray}
and we add six free parameters $\mathrm{A_{i=1,...6}}$ to model the 
amplitudes of the CIB shot-noise power spectra. All parameters have a 
uniform prior, and we fix the emissivity index to $\mathrm{\beta=1.5}$ 
\citep{planck_dust2014}, and $\sigma^2_{L/M}=0.5$ \citep{shang2012,planckXXX}.
\begin{table*}
\centering
\begin{tabular}{|c|c|c|c|c|c|c|}
\cline{2-2}
\hline
Line & A & $\sigma_A$ & B & $\sigma_B$ & Transition & Temperature (K)   \\
$\mathrm{[OI]}$ $63.2$ $\mu$m & $0.98$ & $0.03$ & $2.70$ & $0.10$ & $\mathrm{^3P_{1}\rightarrow ^3P_{2}}$ & $228$\\
$\mathrm{[NII]}$ $121.9$ $\mu$m & $1.01$ & $0.04$ & $3.54$ & $0.11$ & $\mathrm{^3P_2\rightarrow ^3P_1}$ & $188$ \\
$\mathrm{[OI]}$ $145.5$ $\mu$m & $0.89$ & $0.06$ & $3.55$ & $0.17$ & $\mathrm{^3P_1\rightarrow ^3P_0}$ & $327$ \\
$\mathrm{[CII]}$ $157.7$ $\mu$m & $0.89$ & $0.03$ & $2.44$ & $0.07$ & $\mathrm{^2P_{3/2}\rightarrow ^2P_{1/2}}$ & $92$\\
$\mathrm{[NII]}$ $205.2$ $\mu$m & $1.01$ & $0.04$ & $4.01$ & $0.11$ & $\mathrm{^3P_1\rightarrow ^3P_0}$ & $70$ \\
\hline
\end{tabular}
\caption{Main parameters to model the luminosity of all emission
lines considered in this paper as a function of the
total infrared luminosity, taken from \cite{spinoglio2012}.
Also shown is the transition level for each line,
with its associated temperature.}
\label{tab:spinoglio}
\end{table*}
With a total $\chi^2$ value of $104.9$ for $97$ degrees of freedom, we
obtain a very good fit to the data. In Table~\ref{tab:mcmc}, we quote
mean values and marginalized limits for all free parameters used in the fit,
while in Fig.~\ref{fig:power_spectrum} we plot the
\textit{Herschel}/SPIRE measurements of the CIB power spectra,
together with our best estimates of the 1-halo, 2-halo, shot-noise,
and total power spectrum. 
%Finally, in Fig~\ref{fig:sfrd} we show
%the data used to constrain the SFRD, {\bf the curve obtained 
%from the best-fit parameters (black line), the best-fit curve obtained 
%from a fit without SFRD data from 
%\cite{madau2014}, except for the value at $\mathrm{z=0.07}$ (green curve), 
%and the SFRD from \cite{planckXXX} (red curve)}.
\begin{table*}
\label{t4}
\centering
\begin{tabular}{|c|c|c|}
%\begin{center}
\cline{2-2}
%     \multicolumn{3}{|c|}{Mean values } \\
%     \multicolumn{3}{|c|}{Mean values and marginalized $68\%$ c.l. for halo model parameters and shot-noise levels (in $\m\athrm{Jy^2}$/sr)} \\ 
 \hline
Parameter & Definition & Mean value \\
     \hline
     $\mathrm{T_d}$ & SED: Redshift-averaged dust temperature & \multicolumn{1}{c|}{$25.3\pm1.1$} \\
     $\mathrm{\delta}$ & Redshift evolution of the normalization of the $\mathrm{L-M}$ relation  & \multicolumn{1}{c|}{$2.6\pm0.2$} \\
     $log(M_{\mathrm{eff}})[\mathrm{M_{\odot}}]$ & Halo model most efficient mass  & \multicolumn{1}{c|}{$12.6\pm0.1$} \\
     $S^{250x250}$ & Shot noise for 250x250 $\mu$m & $<7237$ ($95$ c.l.) \\
     $S^{250x350}$ & Shot noise for 250x350 $\mu$m & $5331\pm151$  \\
     $S^{250x500}$ & Shot noise for 250x500 $\mu$m & $2806\pm93$  \\
     $S^{350x350}$ & Shot noise for 350x350 $\mu$m & $4677\pm124$ \\
     $S^{350x500}$ & Shot noise for 350x500 $\mu$m & $2659\pm80$ \\
     $S^{500x500}$ & Shot noise for 500x500 $\mu$m & $1600\pm61$ \\
     \hline
     \end{tabular}
\caption{\label{tab:mcmc} Mean values and, where not otherwise stated, marginalized $68\%$ c.l. 
for halo model parameters and shot-noise levels (in $\mathrm{Jy^2}$/sr) from the MCMC fit 
using \textit{Herschel}/SPIRE measurements.}
\end{table*}

\begin{figure}[!t]
\begin{center}
\includegraphics[width=90mm]{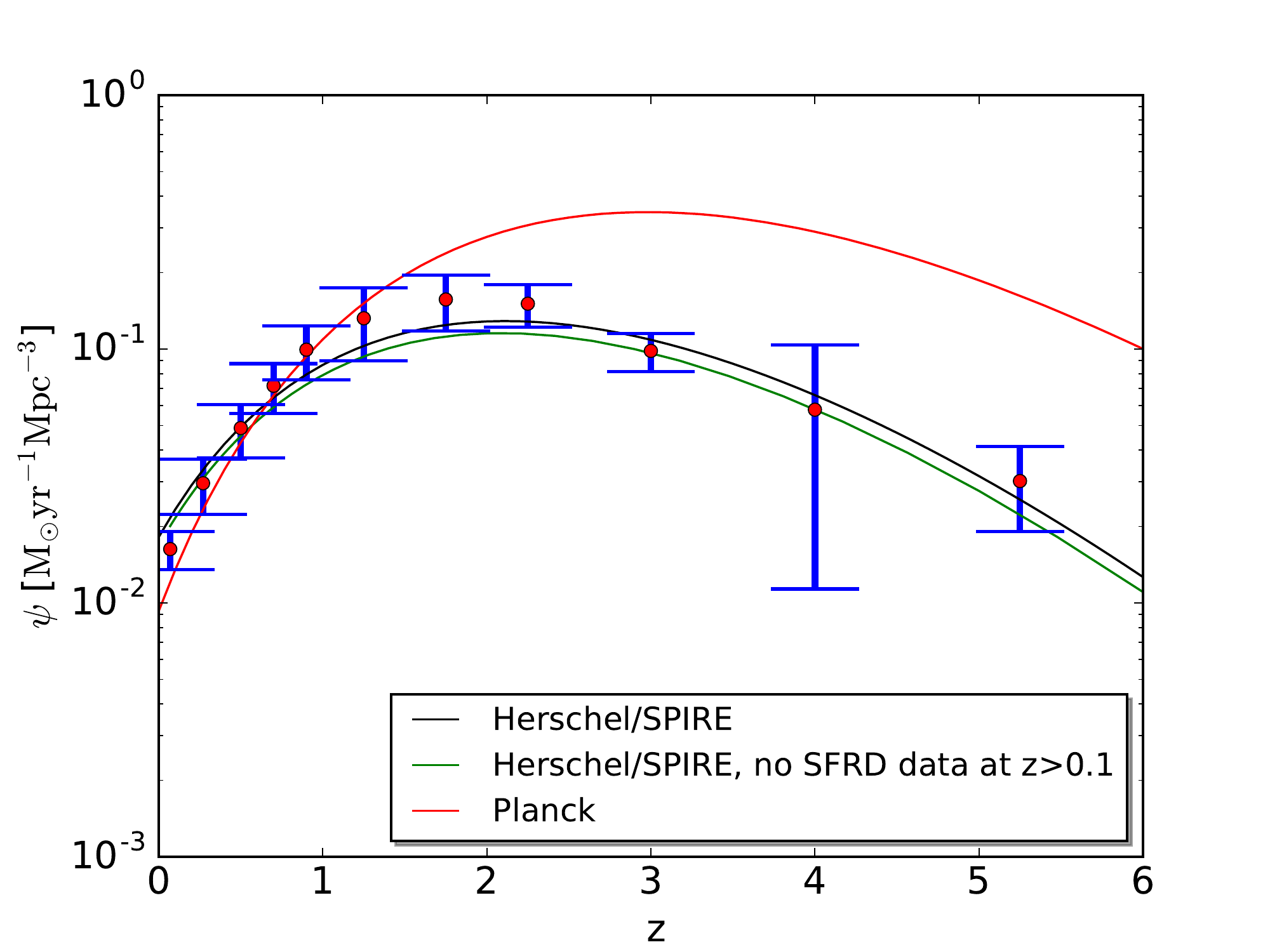}
\caption{Best fit estimates of the SFRD using {\it Herschel}/SPIRE 
CIB clustering measurements combined with a 
compilation of data extracted from \cite{madau2014} either 
in the range $\mathrm{0<z<6}$ (black line), or in $\mathrm{0<z<0.1}$ 
(one single measurement at $\mathrm{z=0.07}$, green line). Also plotted is the estimate from \cite{planckXXX} (red line).}
\label{fig:sfrd}
\end{center}
\end{figure}
\begin{figure}[!t]
\begin{center}
\includegraphics[width=90mm]{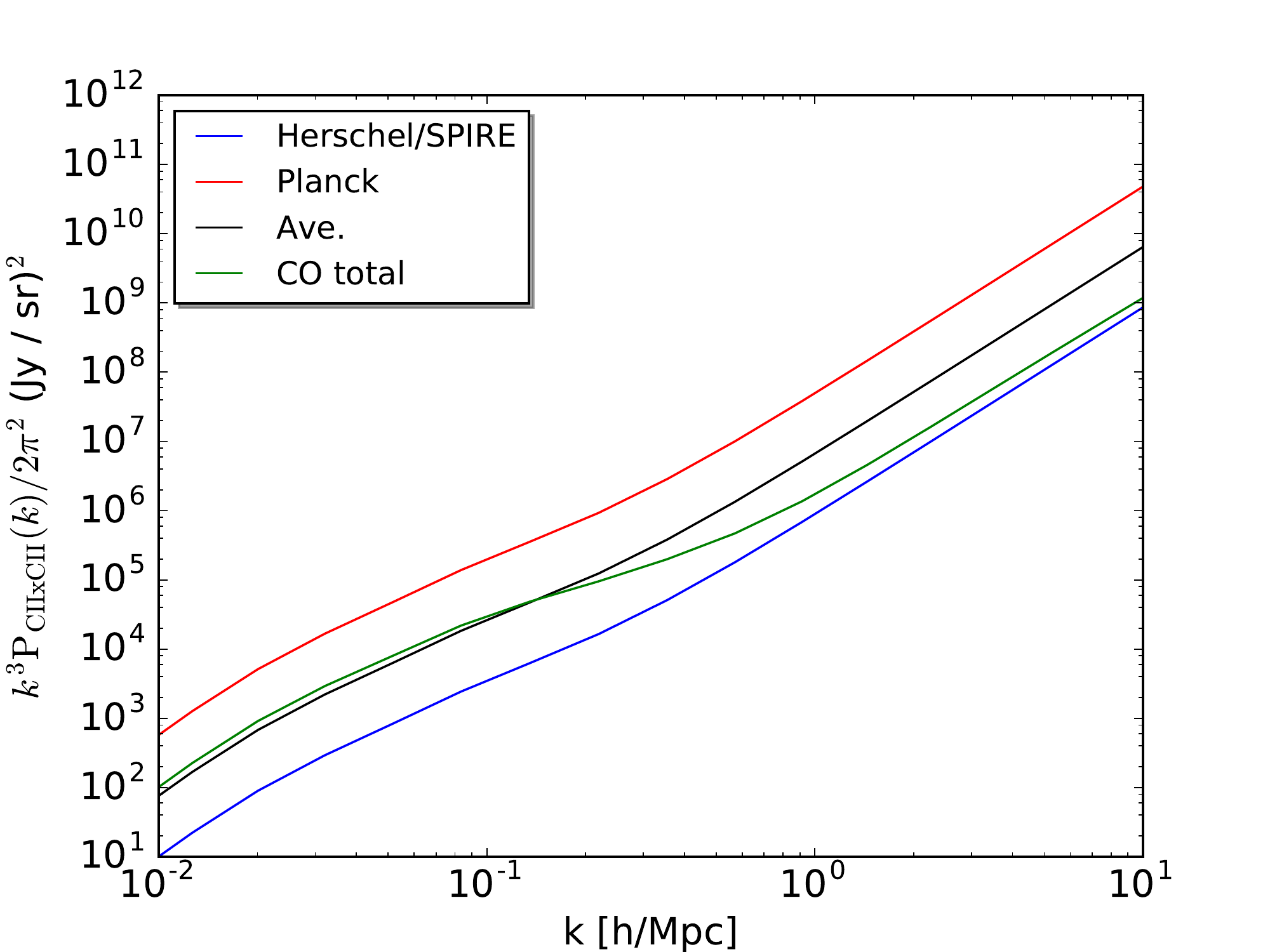}
\caption{Average estimate of the [CII] auto-power spectrum at 
redshift z=7 (black line), together with an optimistic estimate 
(red line) obtained from the mean values of the 
Halo model parameters from \cite{planckXXX}, and an estimate (blue line) 
from our analysis of {\it Herschel}/SPIRE data. Also plotted is the CO power spectrum 
computed as the sum of the transitions from CO(3-2) to CO(7-6).}
\label{fig:best_power}
\end{center}
\end{figure}

It is important to note that there is a relevant uncertainty 
associated to measurements of the SFRD, 
especially at the high redshifts considered in this work. 
The compilation of measurements extrapolated 
from \cite{madau2014} (plotted in Fig.~\ref{fig:sfrd}), 
is based on galaxy counts, and there is a number of 
uncertain steps in the conversion from 
galaxy counts and luminosities to star formation rates, 
mainly related to assumptions on
conversion factors and dust attenuation. 
When considering clustering meausurements, 
the Planck Collaboration, using a Halo model 
similar to the one presented in this paper, 
and fitting to CIB power spectra 
between 217 GHz (1381 $\mu$m) and 857 GHz (350 $\mu$m) 
in the multipole range $\mathrm{50<l<2000}$, infer a 
much higher SFRD at high redshifts \citep{planckXXX}, 
respect to the values found here by fitting Herschel-SPIRE data and 
star formation rate density data from \cite{madau2014} 
(see also discussion in \cite{cheng2016}). 
Similar results have been obtained by cross-correlating 
the CIB with the CMB lensing (\cite{planckXVIII}, 
see also Fig.~14 of \cite{planckXXX}).
The reason for this discrepancy is mainly
due to the different values inferred for 
the parameter $\mathrm{\delta}$ in 
Eq.~\ref{eqn:phiz}. \cite{planckXXX} found 
$\mathrm{\delta = 3.6\pm0.2}$ (see Table 9 of \cite{planckXXX}),
while we find $\mathrm{\delta=2.6\pm0.2}$, 
compatible with \citep{viero2013a}. 
We checked that the fitting to 
star formation rate density data from \cite{madau2014} is not 
responsible for such a divergence, 
by performing an MCMC run with only one measurement of the
local SFRD at $\mathrm{z=0.07}$ from \cite{madau2014} 
(thus being compatible with {\it Planck}'s analysis,
since they use a prior on the local SFRD from \cite{vaccari2010}).
As it is clear from Fig.~\ref{fig:sfrd}, we are not able
to obtain SFRD values compatible
with \cite{planckXXX} at high redshifts.\\
The disagreement between our analysis and results from 
\cite{planckXXX} can be explained by a combination 
of multiple factors involving our ignorance of the 
exact values of some key parameters, 
such as the amplitudes of the shot noise power spectra 
and the redshift evolution of the galaxy luminosity, 
coupled to differences in the datasets considered. 
CIB anisotropies are mostly sourced by galaxies 
at redshift $\mathrm{1<z<4}$ and, 
in this range, a simple power law might not be a good description of 
the redshift evolution of the
galaxy luminosity/halo mass relation. Some semianalytic models of galaxy
formation and evolution find a power law slope of $\mathrm{\sim2.5}$
\citep{delucia2007,neistein2008}, but also a
more gradual evolution, with different slopes for
low redshift and high redshift sources \citep{wu2016}.
On the other end, observations are more in agreement with a steep
evolution with redshift \citep{oliver2010}, or with a steep evolution
followed by a plateau for $\mathrm{z\sim2}$
\citep{bouche2010,weinmann2011}, which is also not easily
explained by theoretical arguments
\citep{bouche2010,weinmann2011}. \cite{planckXXX} is indeed
able to find lower values for the star
formation rate density at early times, 
more in agreement with this work, but 
only when they impose the condition $\mathrm{\delta=0}$ for
$\mathrm{z\ge2}$ (see Fig.~14 of \cite{planckXXX}).\\
The differences between the two datasets in terms of 
angular scales and related uncertainties 
can also be responsible for the difference values 
inferred for $\mathrm{\delta}$. {\it Planck} 
data probe CIB anisotropies at large scales 
with very high precision. However, because of its 
angular resolution, {\it Planck} is not able to access 
multipoles higher than $\mathrm{l\sim2000}$,
where the 1-halo term and the shot-noise dominate the clustering,
and are degenerate. Uncertainties in the 
contribution of these two terms to the small-scale 
clustering (\cite{planckXXX} used free amplitudes 
for the shot-noise power spectra, with flat 
priors based on current measurements, such as, e.g., \cite{bethermin2012model}) 
translates in an uncertainty in the inferred
constraints on the halo model parameters.
On the other end, {\it Herschel}/SPIRE data probe
both large and small scales, 
but while adding information at small scales helps 
disentangling the relative contributions to the total power from the 
1-halo term and the shot-noise, 
the largest scales are measured with much larger uncertainty 
than {\it Planck}. Finally, {\it Planck} and {\it Herschel} probe a different 
frequency range, which might affects results. Thus, it is possible that the
differences in the datasets used, coupled with uncertainties regarding the 
level of the shot-noises, and a poor 
description of the redshift evolution of the sources, determine 
different values for the parameter $\mathrm{\delta}$.\\
It is clear that the higher the value of the star 
formation rate density, the greater the value for the 
mean emission from all atoms and molecules. 
This would translate in large amplitude for the emission line power spectra. 
In order to be as independent as possible on 
the particular values of the Halo model parameters used to constrain the galaxy infrared 
luminosity, we compute predictions for the 3D power spectra of emission 
lines using both the mean values found 
by fitting {\it Herschel}/SPIRE data (quoted in Table~\ref{tab:mcmc}) and the 
mean values quoted in Table~9 of \cite{planckXXX}. 
The geometric average of these two estimates will be 
our best estimate of the power spectrum 
of the emission lines. In the rest of the paper we will focus on 
predictions based on these average estimates of the power spectra. In 
Fig.~\ref{fig:best_power} we show the 3D power spectrum of 
[CII] emission at redshift $\mathrm{z=7}$ 
obtained by using mean parameter values for the Halo 
model parameters from \cite{planckXXX} (optimistic scenario), 
mean parameter values from our 
analysis of {\it Herschel}/SPIRE data, and their average. The ``average'' model 
considered here agrees at both large and 
small scales with the model prediction from \cite{gong2011b}, which is  
based on a physical model that takes into account for the spontaneous, 
stimulated and collisional emission to compute the CII spin temperature. However, it 
predicts shot-noise amplitudes higher than what found in \cite{silva2015,lidz2016}.

\subsection{Intensity mapping power spectrum from the Halo model}
\label{sect:3dImapping}
The analysis presented in the previous section has been necessary to 
constrain the main parameters describing the galaxy SED and its dependence
on halo mass and redshift.\\
The galaxy infrared luminosity is:
\begin{eqnarray}
L_{\mathrm{IR}} &=& \int_{\mathrm{300\,GHz}}^{\mathrm{37.5\,THz}}\,\Theta[(1+z)\nu]d\nu
\end{eqnarray}
where the extremes of integration correspond to the wavelength range
$\mathrm{8<\lambda<1000\,\mu}$m. We can use scaling relations 
provided in \cite{spinoglio2012}, to express the emission line 
luminosity $\mathrm{I_{\alpha}}$ 
(where $\alpha$ denotes emission lines from the atoms and molecules 
considered, such as carbon, oxygen, nitrogen) 
as a function of the constrained infrared luminosity, as:
\begin{eqnarray}
\label{eqn:LIR}
log_{10}(I_{\alpha}) &=& (A\pm\sigma_A)log_{10}(L_{IR}) - (B\pm\sigma_B),
\end{eqnarray}
where all luminosities are in 
units of $\mathrm{10^{41}}$ erg\,s$^{-1}$. 
These scaling relations are obtained from a sample of local galaxies compiled by 
\cite{brauher2008} using all observations collected 
by the LWS spectrometer \citep{clegg1996} onboard ISO \citep{kessler1996}. 
Regarding the [NII] 205 $\mathrm{\mu}$m emission line, whose luminosity is not 
found in \cite{spinoglio2012}, we assume that it is three times 
weaker than the [NII] 122 $\mathrm{\mu}$m; this values is in 
agreement with both theoretical expectations and 
recent measurements \citep{oberst2011,zhao2016}, although it is 
higher than what recently found in our Galaxy \citep{goldsmith2015}.\\
In Table~\ref{tab:spinoglio} we summarize the values used  
for slopes, intercepts and their uncertainties, together with 
their associated transition, and transition 
temperatures from \cite{kaufman1999,cormier2015}.\\
The emission line luminosity at each redshift for each halo mass 
can now be expressed as previosly done for the galaxy luminosity 
(see Eq.~\ref{eqn:luminosity}) as:
\begin{eqnarray}
L_{\alpha}(M,z) &=& F(M,z)I_{\alpha},
\end{eqnarray}
where the term $\mathrm{F(M,z)}$ contains the global dependence on 
redshift and halo mass as
\begin{eqnarray}
F(M,z) &=& L_0\Phi(z)\Sigma(M),
\end{eqnarray}
and we use the parameter values from Table~\ref{tab:mcmc} 
to compute the term $\mathrm{F(M,z)}$. This functional form allows us 
to link the emission line luminosity of a galaxy to its host halo mass, 
and to evolve the amplitude of all emission lines with redshifts. 
We note that this model assumes that the redshift 
evolution of all emission lines is the same, since 
it follows the evolution of the galaxy infrared luminosity (through the 
parameter $\mathrm{\delta}$). Different emission lines might 
have different a evolution with redshift, and more sophisticated models 
could incorporate redshift-dependent scaling relations for each line. However, 
current data do not allow us to constrain the exact dependence on redshift 
of each emission lines. Thus, to keep the analysis as simple as possible, 
we do not consider such a scenario.\\
It is easy to see that, assuming that each halo hosts only one galaxy 
(a good approximation because, at high redshift, halos are not very massive, 
see also \cite{lidz2011}), and in the limit of sufficiently large scales (so that the NFW profile 
approaches unity), 
the clustering auto-power spectrum of emission line $\alpha$ can
be written as:
\begin{eqnarray}
\label{eqn:em_eq2}
P_{\alpha\alpha}(k,z) &=& K_{\alpha}^2(k,z)P_{\mathrm{lin}}(k,z),
\end{eqnarray}
where
\begin{eqnarray}
K_{\alpha}(k,z)&=&\int_{M_{\mathrm{min}}}^{\infty}dM
\frac{dN}{dM}b(M,z)\frac{L_{\mathrm{\alpha}}(M,z)}{4\pi}.\\
\nonumber
\end{eqnarray}
Introducing an effective, scale independent, bias term as:
\begin{eqnarray}
b_{\mathrm{eff}}(z) &=& \frac{\int\,dM\frac{dN}{dM}b(M,z)\Sigma(M)}{\int\,dM\frac{dN}{dM}\Sigma(M)}
\label{eqn:eff_bias}
\end{eqnarray}
the clustering power spectrum of emission line $\alpha$ can be expressed as:
\begin{eqnarray}
P^{\mathrm{clust}}_{\alpha\alpha}(k,z) &=& b_{\mathrm{eff}}^2(z)\bar{I}^2_{\alpha}(z)P_{lin}(k,z),
\label{eqn:IM_clust}
\end{eqnarray}
where the average specific intensity $\mathrm{\bar{I}_{\alpha}(z)}$ is:
\begin{eqnarray}
\bar{I}_{\alpha}(z) = \frac{1}{4\pi}\frac{c}{H(z_{\alpha})}\frac{1}{\nu_{\alpha}}\int\,dM\frac{dN}{dM}L_{\alpha}(M,z),
\end{eqnarray}
and $z_{\alpha}$ denotes the redshift of emission of the atom or molecule $\alpha$.
Analogously, the shot-noise power spectrum can be expressed as:
\begin{eqnarray}
P^{\mathrm{SN}}_{\alpha\alpha} &=& 
\bar{I}_{\alpha}^2(z)\frac{\int\,dM\frac{dN}{dM}\Sigma(M)^2}{\Big(\int\,dM\frac{dN}{dM}\Sigma(M)\Big)^2}. 
\label{eqn:IM_SN}
\end{eqnarray}

\section{The physics of the ISM with emission lines and emission line ratios}
\label{sect:ism}
Understanding the main heating and cooling processes of the ISM is a key goal 
of astronomy, because they play a fundamental role in the formation 
of stars, and thus in the galaxy evolution.
Space missions such as {\it Planck} and {\it Herschel}, together with the 
Stratospheric Observatory for Infrared Astronomy (SOFIA) and the Atacama 
Large Millimeter Array (ALMA), are now 
giving new insights on these physical processes, providing 
spatially resolved maps of the interstellar dust in our Galaxy, 
and measuring atomic and molecular emission lines from 
the main phases of the ISM both in the 
Milky Way \citep{pineda2013,pineda2014,goicoechea2015}, and in 
external galaxies (see e.g. \cite{stacey2010,scoville2014,capak2015,gullver2015,blain2015,bethermin2016,aravena2016}).\\
The gas in the ISM of galaxies is observed in three main phases; 
a cold and dense neutral medium (T$\mathrm{\geq50}$K) 
is in rough pressure equilibrium 
(with $\mathrm{P/k\sim10^3-10^4}$Kcm$^{-3}$) 
with a hot (T$\mathrm{\geq10^6}$K), ionized phase, 
and an intermediate, warm (T$\mathrm{\geq\,8000}$K) phase, 
which can be either neutral or ionized, depending on the gas density 
\citep{wolfire1995}.\\
Various mechanisms contribute to the heating and cooling of the ISM. 
For a gas with hydrogen density n, temperature T, cooling 
rate per unit volume of $\mathrm{\Lambda(T)}$, 
and heating rate per unit volume of $\mathrm{\Gamma(T)}$, 
the thermal balance between heating and 
cooling is expressed in terms of a Generalized Loss Function L:
\begin{eqnarray}
L(n,T) &=& \Lambda(T) - \Gamma(T).
\end{eqnarray}
For a gas at constant thermal pressure nT, 
equilibrium occurs when $\mathrm{L=0}$ and the explicit 
form for $\mathrm{\Lambda}$ and $\mathrm{\Gamma}$ depends 
on the heating and cooling process
considered, as explained below.\\
The investigation of the thermal balance and stability conditions of the 
neutral ISM started with \cite{field1969}, who first presented 
a model of the ISM based on two thermally stable neutral phases, 
cold and warm, heated by cosmic-rays. 
Subsequent analyses by many authors focused on the heating 
provided by the photoelectric ejection of electrons from dust 
grains by the interstellar radiation field 
\citep{draine1978,wolfire1995,kaufman1999,wolfire2003}.  
Most of the Far-Ultraviolet (FUV) 
starlight impinging on the cold neutral medium  
is absorbed by dust and large molecules of 
polycyclic aromatic hydrocarbons (PAH), and then reradiated 
as PAH infrared lines and infrared continuum
radiation. However, as pointed out by \cite{tielens1985a}, 
in photodissociation regions (PDRs), 
the photoelectric heating of dust grains 
provides an efficient mechanism 
($0.1\%-1\%$) at converting the FUV heating into
atomic and molecular gaseous line emission. 
The physics of heating processes in PDRs can be understood in 
terms of a limited set of parameters, namely the density of 
Hydrogen nuclei density $\mathrm{n}$ 
and the incident FUV ($6$eV$\,\mathrm{<h\nu<13.6}$ eV) 
parametrized in units of the local interstellar field, 
$\mathrm{G_0}$ \citep{tielens1985b,tielens1985a,kaufman1999}, 
in units of the Habing field ($\mathrm{1.6\cdot10^{-3}}$ ergs cm$\mathrm{^{-2}s^{-1}}$). 
The basic mechanism for 
gas heating and cooling is the following: 
about 10$\%$ of incident FUV photons eject photoelectrons from 
dust grains and PAH molecules, which cool by continuum infrared emission. 
The photoelectrons (with energy of about 1 eV) heat the gas by collisions, 
and the gas subsequently cools via FIR fine-structure line emission. 
The entire process thus results in the conversion 
of FUV photons to FIR continuum emission plus spectral line emission 
from various atoms and molecules. As an example, the computation of the 
heating due to small grains is given by \citep{bakes1994}:
\begin{eqnarray}
\Gamma &=& 10^{-24}\epsilon\,G_0n_H \,\,\,\,\, \mathrm{erg\,cm^{-3}s^{-1}};
\end{eqnarray}
the radiation field $\mathrm{G_0}$ quantifies the starlight intensity, and 
$\epsilon$ is the fraction of FUV photons absorbed by grains 
which is converted to gas heating (i.e. heating efficiency), 
and it depends on $\mathrm{G_0T^{1/2}n_e}$, 
where $\mathrm{n_e}$ denotes the electron density \cite{wolfire1995}.   
A detailed calculation of the main heating
processes in the ISM, including the effect from photoelectric heating,
 cosmic rays,
soft X-rays, and photoionization of CI is presented in 
\cite{wolfire1995,Meijerink2005}. \\
The cooling rate $\Lambda$ of each atom/molecule depends on 
both the number density and the equivalent temperature of 
each species. A recent estimate of the cooling rate of the 
[CII] line for temperatures between 20\,K and 400\,K\ is 
\citep{wiesenfeld2014}:
\begin{eqnarray}
\Lambda_{\mathrm{CII}} &=& 10^{-24}\Big(11.5+4.0e^{-100K/T_{\mathrm{kin}}}\Big)\\
\nonumber
& & e^{-91.25K/T_{\mathrm{kin}}}n(C^+)n_{H_2}\,\,\,\,\,\mathrm{erg\,cm^{-3}s^{-1}}
\end{eqnarray}
where $\mathrm{n(C^+)}$ denotes the carbon number density, 
and $\mathrm{T_{kin}}$ the kinetic temperature of the gas.\\
Numerical codes compute a simultaneous solution for 
the chemistry, radiative transfer, and thermal balance of PDRs, 
providing a phenomenological description of the interplay among three 
main parameters n, $\mathrm{G_0}$ and T (see e.g. \cite{kaufman1999}) 
for all emission lines. The observed intensity of line emissions can 
thus be compared with models to constrain these 
parameters.\\
Far-infrared emission lines from forbidden atomic
fine-structure transitions such as [CII] (157.7 $\mathrm{\mu}$m),
[OI] (63 $\mathrm{\mu}$m and 145.5 $\mathrm{\mu}$m), are the main 
coolants of the neutral regions of the ISM, and provide many 
insights on the physics of PDRs. 
Other lines, such as [NII] (122 $\mathrm{\mu}$m and 205 $\mathrm{\mu}$m), 
[OIII] (88 $\mathrm{\mu}$m), and [NIII] (57 $\mathrm{\mu}$m), being emitted only in 
ionized regions, complement the study of the ISM probing a different phase.\\
For ground-based surveys such as Time-PILOT 
\cite{crites2014} or CONCERTO, covering approximately the range $\mathrm{200<\nu<300}$ GHz, 
and targeting high redshift ($5<z<8$) galaxies, emission from [CII], 
[OI] (145 $\mathrm{\mu}$m) and [NII] (122 $\mathrm{\mu}$m and 205 $\mathrm{\mu}$m) are accessible. 
A future space-based survey with 
characteristics similar to PIXIE will be able to 
detect most of the main cooling 
lines from both PDRs and from the ionized medium of high redshift galaxies. 
Below we summarize some useful diagnostics of the ISM provided by these 
important lines (see also \cite{cormier2015}).
\begin{itemize}
\item {\bf [CII] emission line}: It is hard to overestimate 
the importance of the [CII] emission line in constraining 
physical properties of the interstellar 
medium. Because of its low ionization potential, the [CII] 
line arises both from ionized and neutral gas. In PDRs, the 
low gas critical density for collisions with Hydrogen 
and the low excitation temperature for the [CII] 
$\mathrm{^2P_{3/2}-^2P_{1/2}}$ transition 
(only 92 K, see Table~\ref{tab:spinoglio}), 
make $\mathrm{C^+}$ one of the major coolant of the neutral ISM. 
Moreover, since the [CII] line is generally one of the brightest lines in 
star-forming galaxies, it is potentially a very strong indicator of star 
formation rate (SFR) \citep{boselli2002,delooze2011,delooze2014,herrera2015}.
As pointed out in \cite{delooze2011}, the tight correlation 
between [CII] emission and mean star formation activity 
is due either to emission from PDRs in 
the immediate surroundings of star-forming regions, or emission 
associated to the cold ISM, thus invoking the Schmidt law to 
explain the link with star formation. Intensity mapping 
measurements of the mean 
amplitude of the [CII] emission line allows 
us to constrain the global star formation activity 
of the Universe at high redshift.
\item {\bf [NII] (122 $\mathrm{\mu}$m and 205 $\mathrm{\mu}$m) emission lines}: With a 
ionization potential of $\mathrm{14.53}$ eV, ionized Nitrogen 
is only found in the ionized phase of the ISM. The two infrared 
[NII] lines are due to the splitting of the ground state of N$^+$ into 
three fine-structure levels, which are excited mainly by collisions with 
free electrons in HII regions, with critical densities of 
290 cm$^{-3}$ and 44 $^{-3}$ for [NII] (122 $\mathrm{\mu}$m) 
and [NII] (205 $\mathrm{\mu}$m) 
respectively, assuming $\mathrm{T_e=8000}$ K, 
see \cite{herrera2016,hudson2004}. 
Being in the same ionization stage, their ratio directly 
determines the electron density of the ionized gas in HII regions. 
For electron densities $n_e$ larger than $10$ cm$^{-3}$, the 
122/205 $\mathrm{\mu}$m line ratio R$_{\mathrm{122/205}}$ 
increases as a function of $n_e$, 
starting from R$_{\mathrm{122/205}}\sim0.6$ for $\mathrm{n_e\sim10}$ 
cm$^{-3}$, and reaching the value R$_{\mathrm{122/205}}\mathrm{\sim3}$ 
(the value used in this paper) for 
$\mathrm{n_e\sim\,100}$ cm$^{-3}$ \citep{tayal2011,goldsmith2015}.\\
Moreover, combined measurements of line emission from [NII] and [CII] can 
be used to estimate the amount of [CII] emission coming from 
the ionized medium \citep{malhotra2001,oberst2006,decarli2014,hughes2016}. 
Recently \cite{goldsmith2015}, using data from the PACS and HIFI instruments 
onboard {\it Herschel}, estimated that between 1/3 and 1/2 of the 
[CII] emission from sources in the Galactic plane arise 
from the ionized gas. The [NII]/[CII] ratio is also useful to estimate 
the metallicity of a galaxy \citep{nagao2012}.
Finally, the [NII] emission lines, arising from gas ionized by O and 
B type stars, directly constrains the ionizing photon rate, and thus 
the star formation rate \citep{bennett1994,mckee1997}.
\item {\bf Oxygen 63 $\mathrm{\mu}$m and 145 $\mathrm{\mu}$m lines}: Oxygen has a
ionization potential of $\mathrm{13.62}$ eV, just above that of 
hydrogen. The $\mathrm{[OI]}$ (63 $\mathrm{\mu}$m) 
and $\mathrm{[OI]}$ (145 $\mathrm{\mu}$m) line emissions 
come from PDRs and, together with [CII], 
are a major coolant of the ISM. However, because 
their fine structure transitions are excited at 
high temperatures (228\,K and 326\,K respectively, against 91\,K 
of [CII]), and their critical densities are quite high 
($\mathrm{\sim\,5e5\,cm^{-3}}$ and $\mathrm{\sim\,1e5\,cm^{-3}}$ for [OI] 
63 $\mathrm{\mu}$m and [OI] 145 $\mathrm{\mu}$m respectively) 
they contribute significantly to the cooling of the ISM 
only for high FUV fields and/or high densities. The measurement of the mean 
amplitude of the [OI] lines with intensity 
mapping would give us clues regarding the mean value of
the $\mathrm{G_0}$ field and the mean density 
of PDRs at high redshifts \citep{Meijerink2007}.
\end{itemize}
The intensity mapping technique would constrain the mean amplitude 
of multiple emission lines, together with their ratio, 
thus probing mean properties 
(such as mean radiation field, mean electron density in 
HII regions, mean density of various atoms, molecules) at high redshifts.
\section{Multiple cross-correlations constrain the physics of the ISM}
As previously stated, the cross-correlation signal 
between different emission lines coming from the same 
redshift is important not only to avoid contamination from foreground 
lines (assuming that, at the frequencies considered in the 
cross-correlation measurements, foregrounds are not correlated), 
but also to help constraining the mean amplitude of each signal. 
This is particularly true at sufficiently small scales, where the 
SNR is larger. If we assume that all lines 
are emitted by the same objects (a reasonable assumption, 
especially if the emission lines are not distant from each 
other, as in the case of the FIR lines such as 
[CII], [NII] and [OI]), it will be possible to constrain 
the mean amplitudes of emission 
lines $\mathrm{I}_{i=1,...N}$, just by looking at all cross-correlation power spectra\footnote{More generally, with enough measurements at high SNR, 
we could always focus on cross-correlation measurements, without 
even bothering with autocorrelations, 
which are complicated by foreground lines.}. \\
For a survey working in a given frequency range where N lines 
are detected, there are $\mathrm{N(N-1)/2}$ 
cross-correlation measurements to be performed and, assuming 
there is perfect correlation among 
lines, it is sufficient 
that $\mathrm{N\geq3}$ to be able to constrain the mean emission 
from all lines.\\
The chances of detecting auto- and cross-power spectra 
strongly depend on the amplitude of
the spectra, which, as already seen, is very uncertain. 
In the following we will consider predicted measurements 
of multiple combinations of emission line power spectra for 
two different surveys. The first one corresponds
to a survey of the [CII] emission line similar 
to the proposed CONCERTO.
The second one, referred in literature as 
CII-Stage II, and described in \cite{silva2015,lidz2016}, 
is more sensitive, and corresponds to an evolution 
of currently planned [CII] surveys.\\
As already emphasized, emission line power spectra are strongly 
contaminated by interlopers lines emitted by molecules at different 
redshifts. In case of [CII], the main confusion results from 
foreground emission of CO molecules undergoing rotational transitions 
between states J and J-1. As an example, [CII] emission from 
$\mathrm{z=6}$ is observed at frequency $\mathrm{\nu_{obs}=271.6}$ GHz, and 
it is mainly contaminated by CO rotational transitions 
$\mathrm{J=3\rightarrow2}$ ($\mathrm{z=0.27}$), 
$\mathrm{J=4\rightarrow3}$ ($\mathrm{z=0.70}$), 
$\mathrm{J=5\rightarrow4}$ ($\mathrm{z=1.12}$), 
$\mathrm{J=6\rightarrow5}$ ($\mathrm{z=1.54}$), and 
$\mathrm{J=7\rightarrow6}$ ($\mathrm{z=1.97}$). Emission lines beyond this 
transition have a negligible contribution to the total foreground due to CO molecules, and we will not consider them in the rest of the paper.\\
Using linear scaling relations from \cite{visbal2010} 
to express the amplitude of the various CO emission lines 
as a function of the infrared luminosity, 
it is possible to estimate the contamination due to the main 
CO rotational lines. In the following, when plotting the 
[CII] auto-power spectra at various redshits, we will also plot 
the CO auto-power spectrum computed as  
the sum of the the main CO rotational transitions involved 
(from $\mathrm{3\rightarrow2}$ to $\mathrm{7\rightarrow6}$), in order to 
highlight the amplitude of this foreground.
\section{Experimental setups and predictions}
\label{sect:exp}
In order to measure high-redshift fluctuations 
with sufficient SNR at the scales of interest, it is important
to optimize the survey area.
All predictions considered in this section are based on measurements spanning
a redshift range $\mathrm{\Delta\,z\sim0.6}$
which corresponds to a frequency range of $\mathrm{B_{\nu}}\sim20$ GHz at 
$z=7$ for the [CII] line. We follow \cite{gong2011b} to compute uncertainties on 
the power spectra.
\begin{figure*}[!t]
\begin{center}$
\begin{array}{ccc}
\includegraphics[width=65mm]{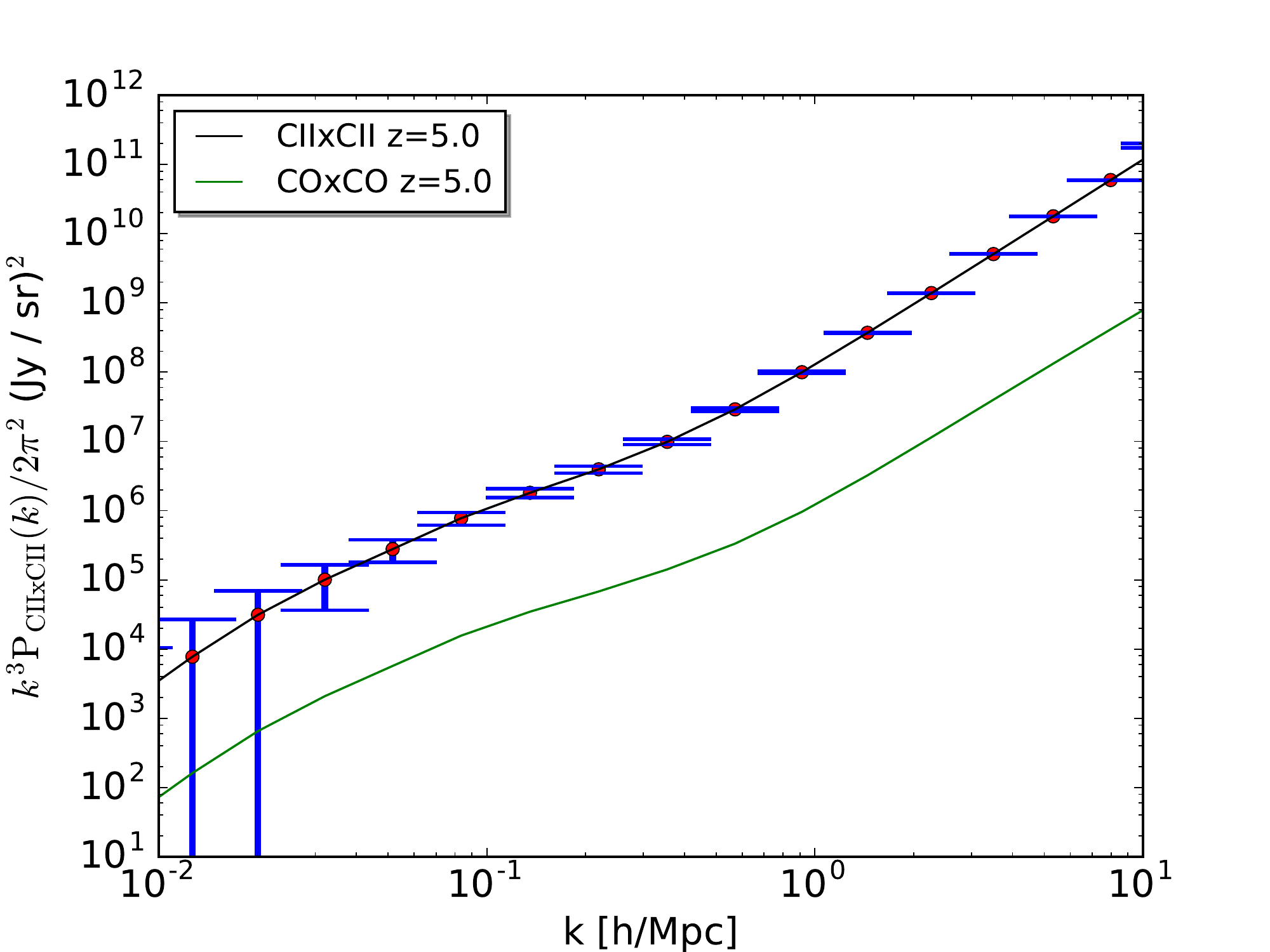}
\includegraphics[width=65mm]{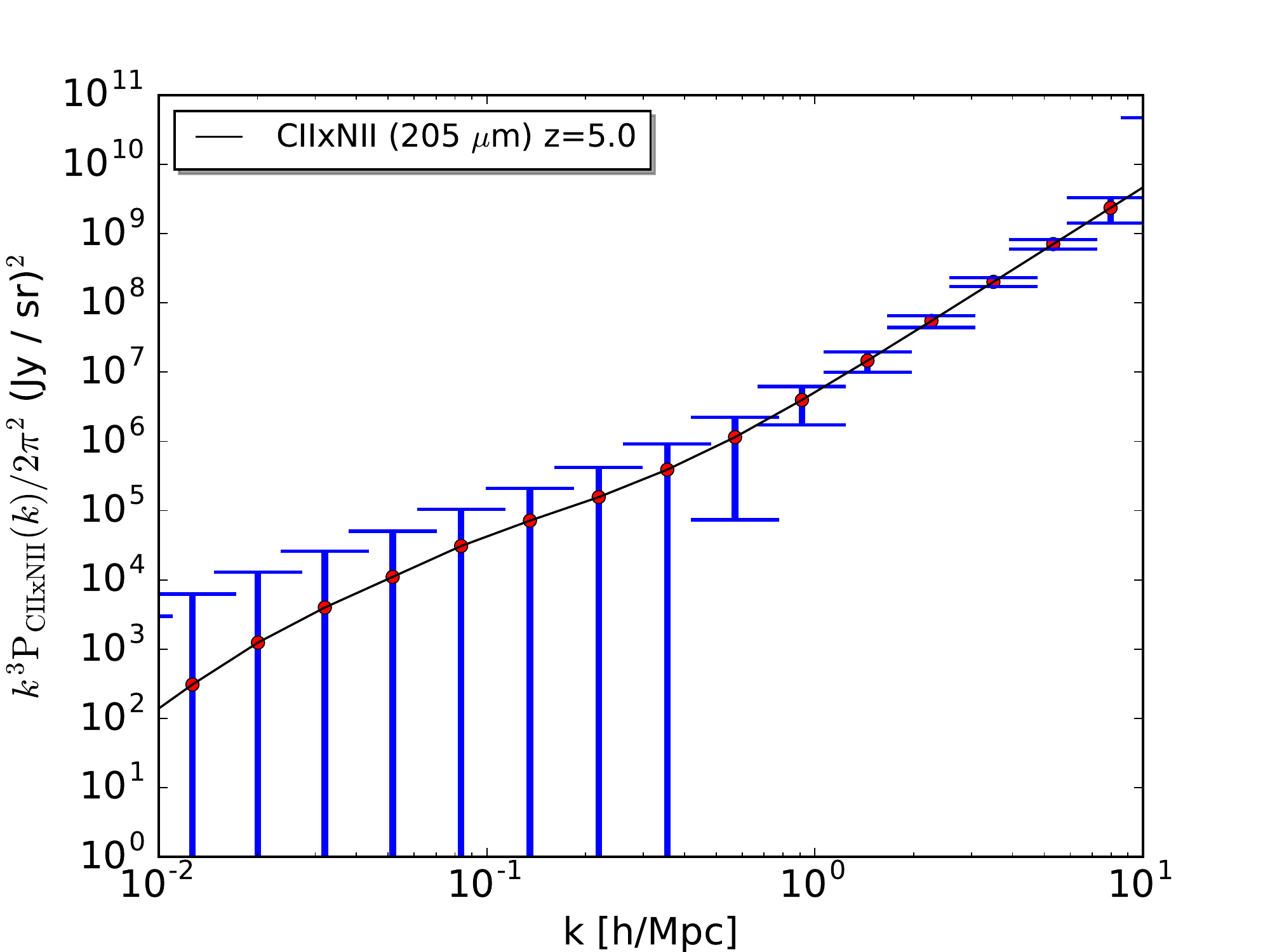}
\includegraphics[width=65mm]{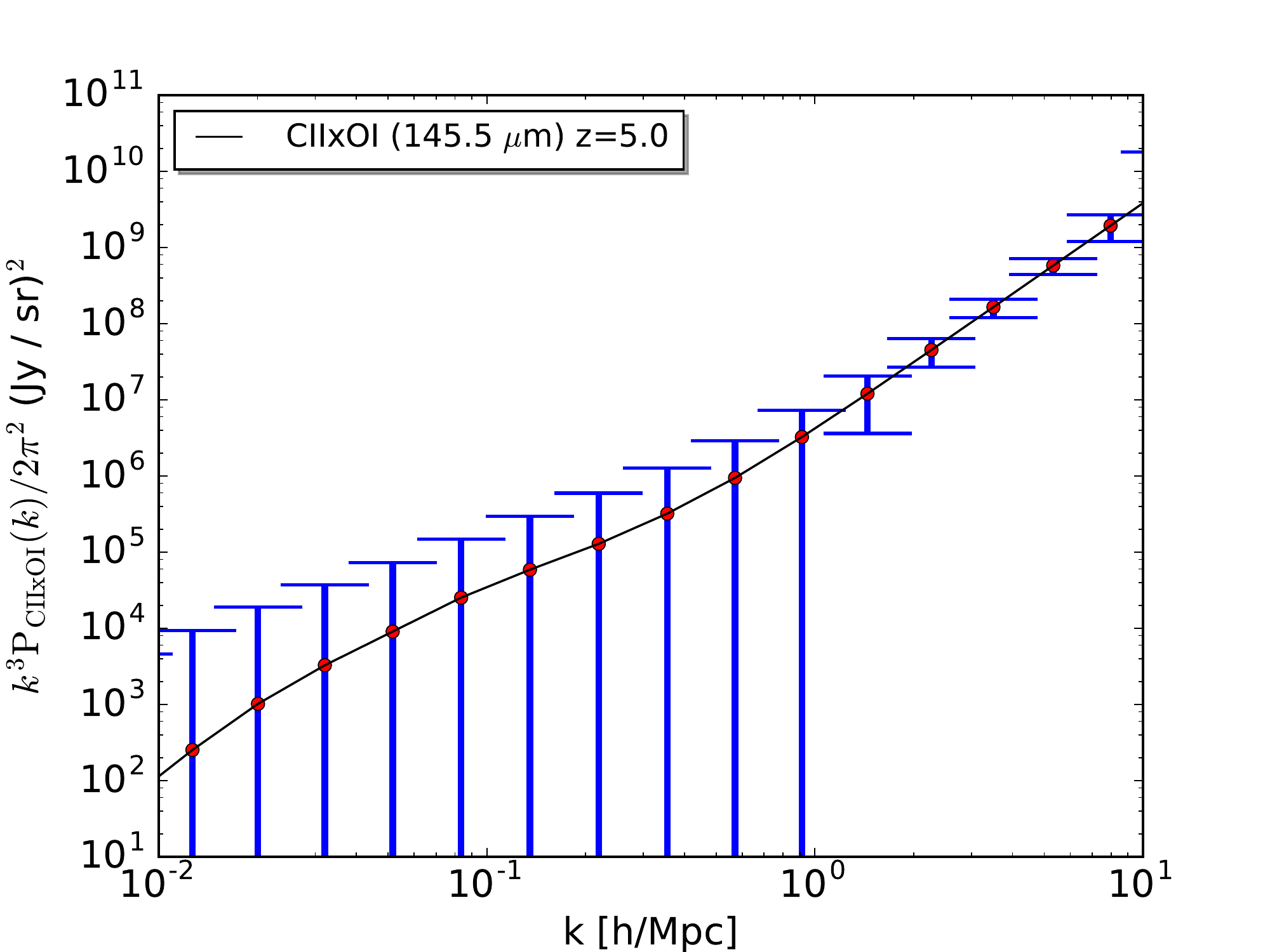} 
\end{array} $
\end{center}
\caption{Predicted [CII] auto-power spectrum and cross-power spectra 
between [CII] and [NII] (205.2 $\mathrm{\mu}$m), and
[OI](145.5 $\mathrm{\mu}$m), at $\mathrm{z=5}$ computed
for the survey CONCERTO. Also plotted in the left panel (green line) is the total CO 
power spectrum computed as the sum of the contributions from CO(3-2) to CO(7-6).}
\label{fig:spectra_concerto_z5}
\end{figure*}

\begin{figure*}[!t]
\begin{center}$
\begin{array}{cc}
\includegraphics[width=85mm]{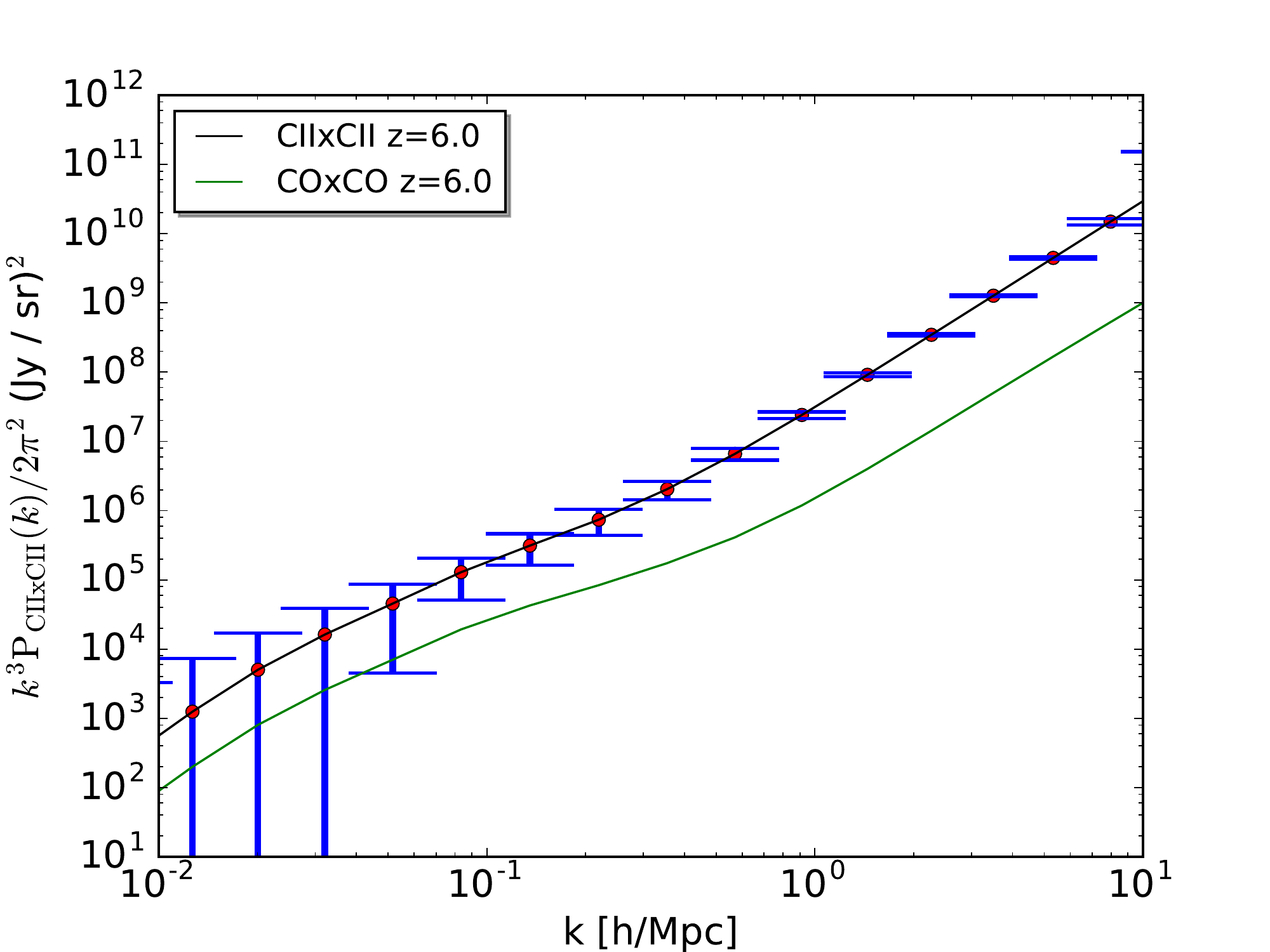}
\includegraphics[width=85mm]{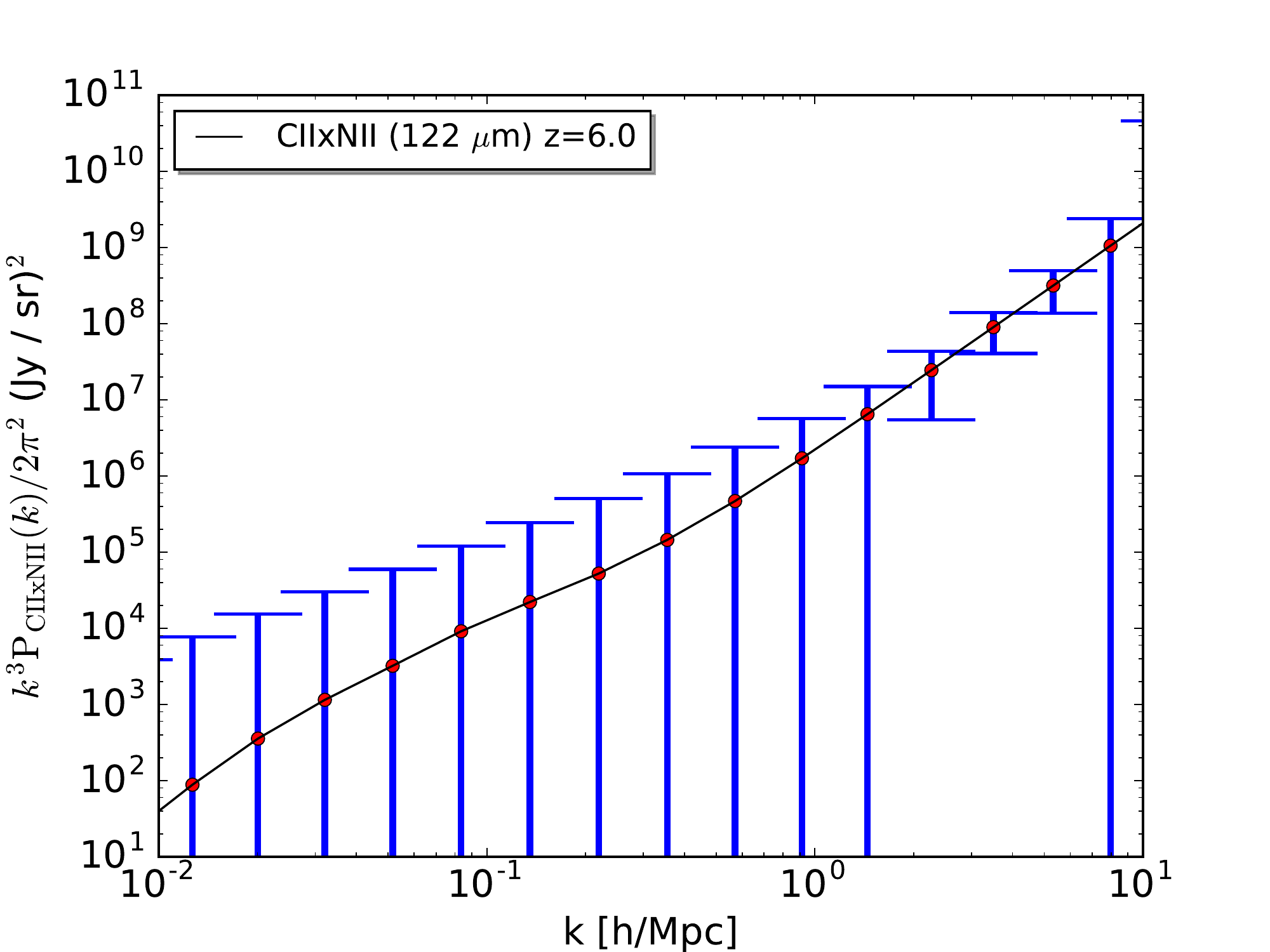}\\
\includegraphics[width=85mm]{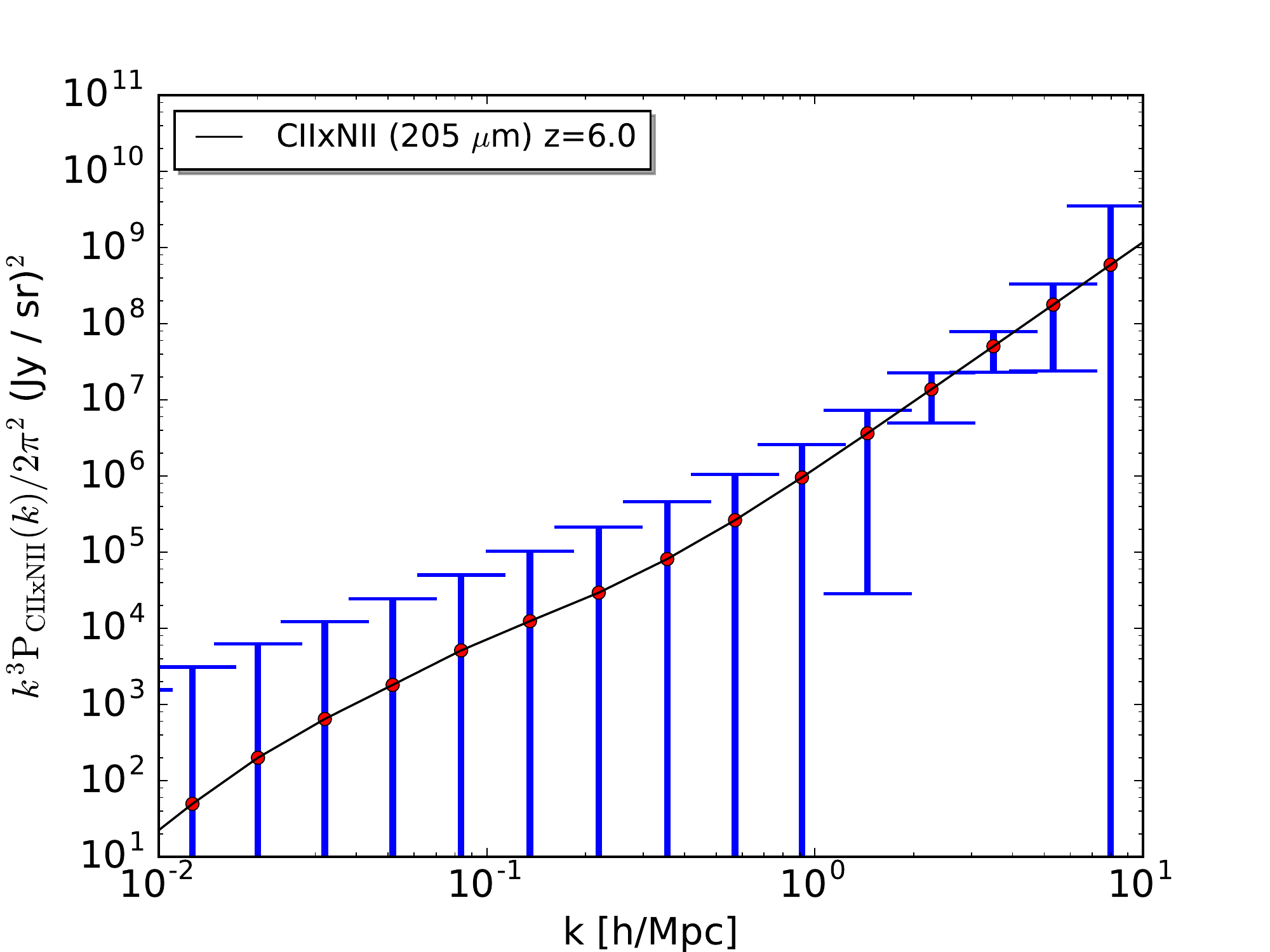}
\includegraphics[width=85mm]{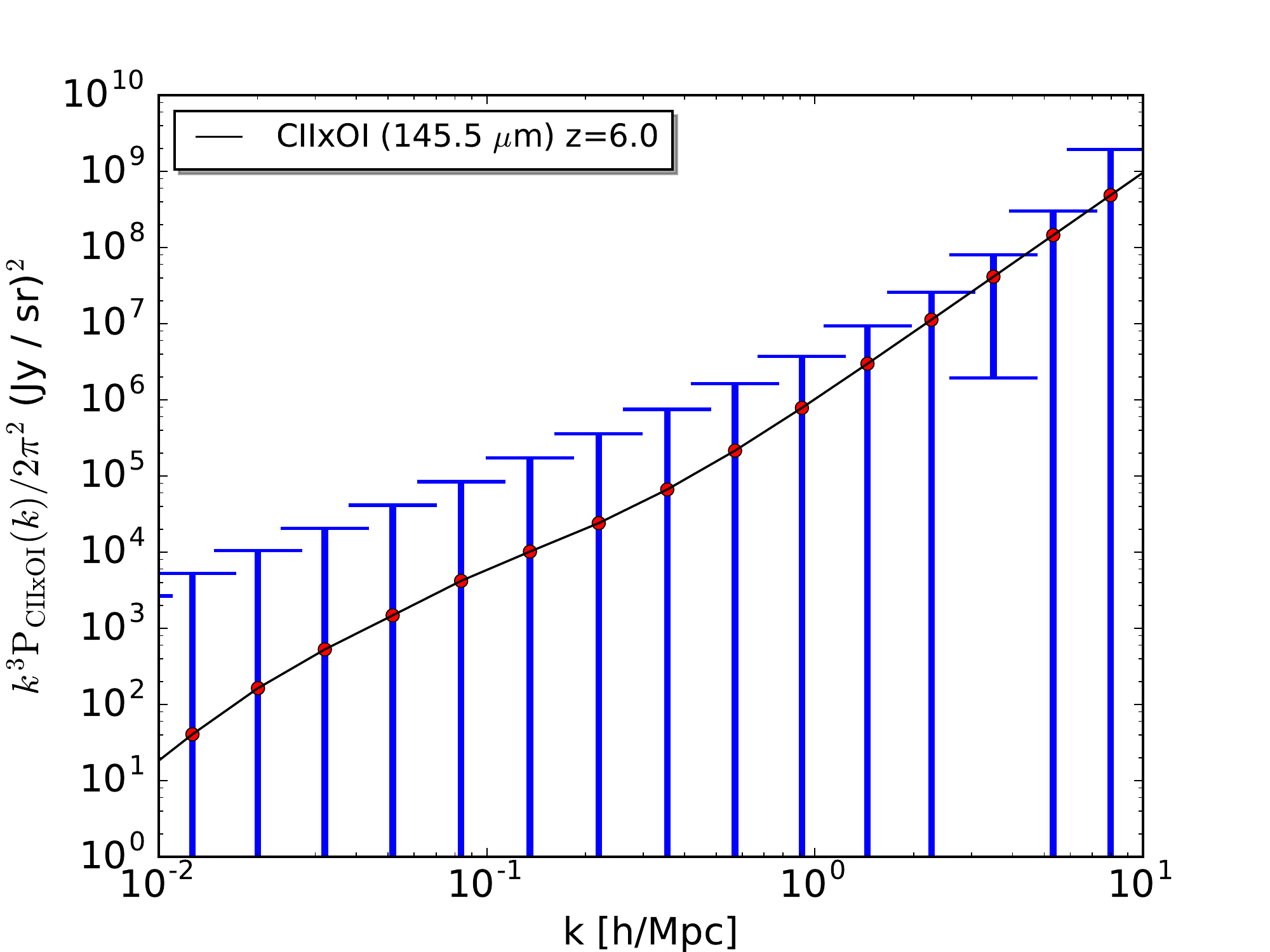}
\end{array} $
\end{center}
\caption{Predicted [CII] and total CO (3-2 to 7-6) auto-power spectra
(left panel, black and green line respectively) at redshift $\mathrm{z=6.0}$, 
and cross-spectra [CII]x[NII] (121.9 $\mathrm{\mu}$m),
[CII]x[NII](205.2 $\mathrm{\mu}$m), and [CII]x[OI](145.5 $\mathrm{\mu}$m) at z=6 for 
the survey CONCERTO.}
\label{fig:spectra_concerto_z6}
\end{figure*}

\begin{figure*}[!t]
\begin{center}$
\begin{array}{ccc}
\includegraphics[width=65mm]{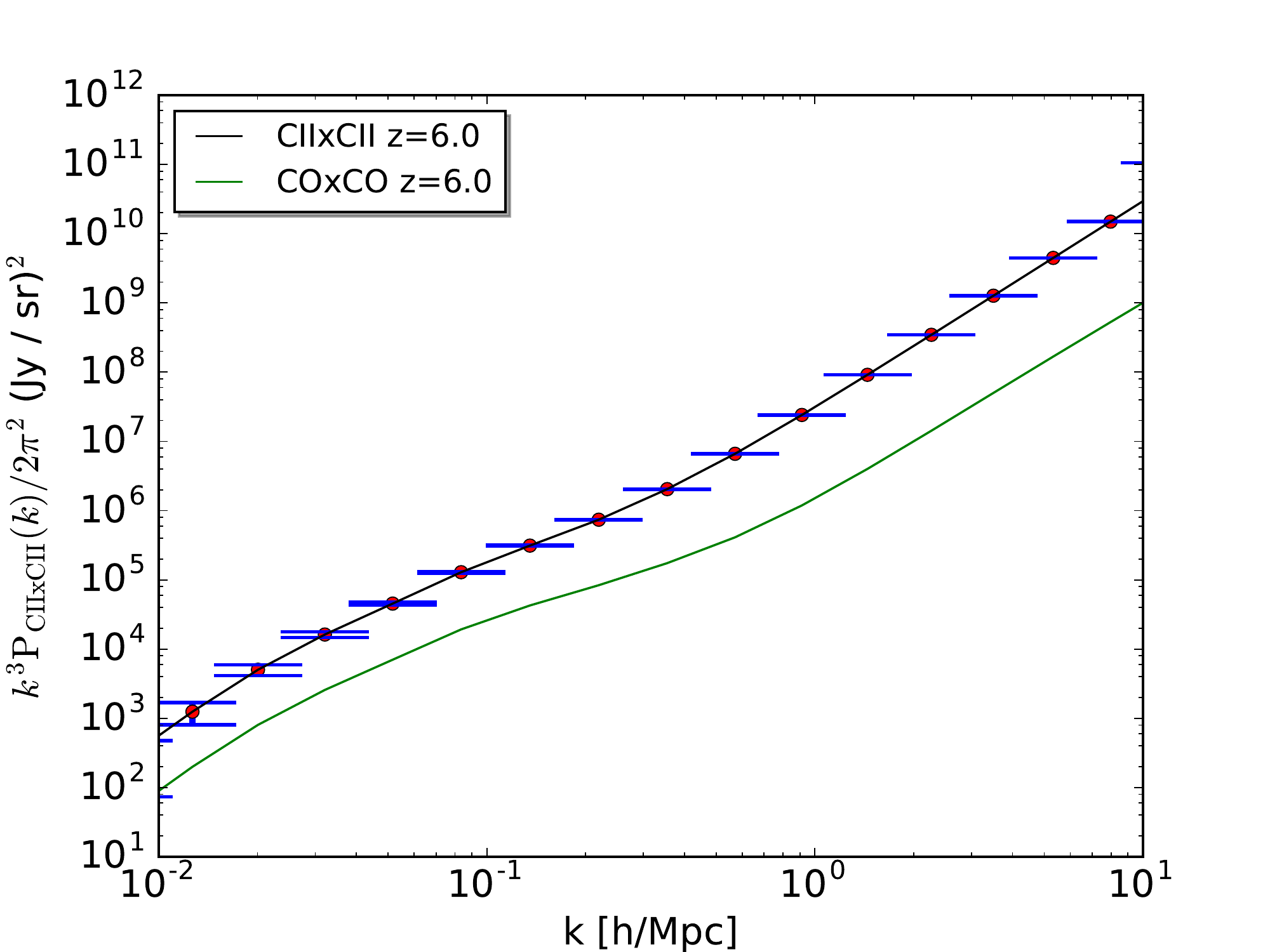}
\includegraphics[width=65mm]{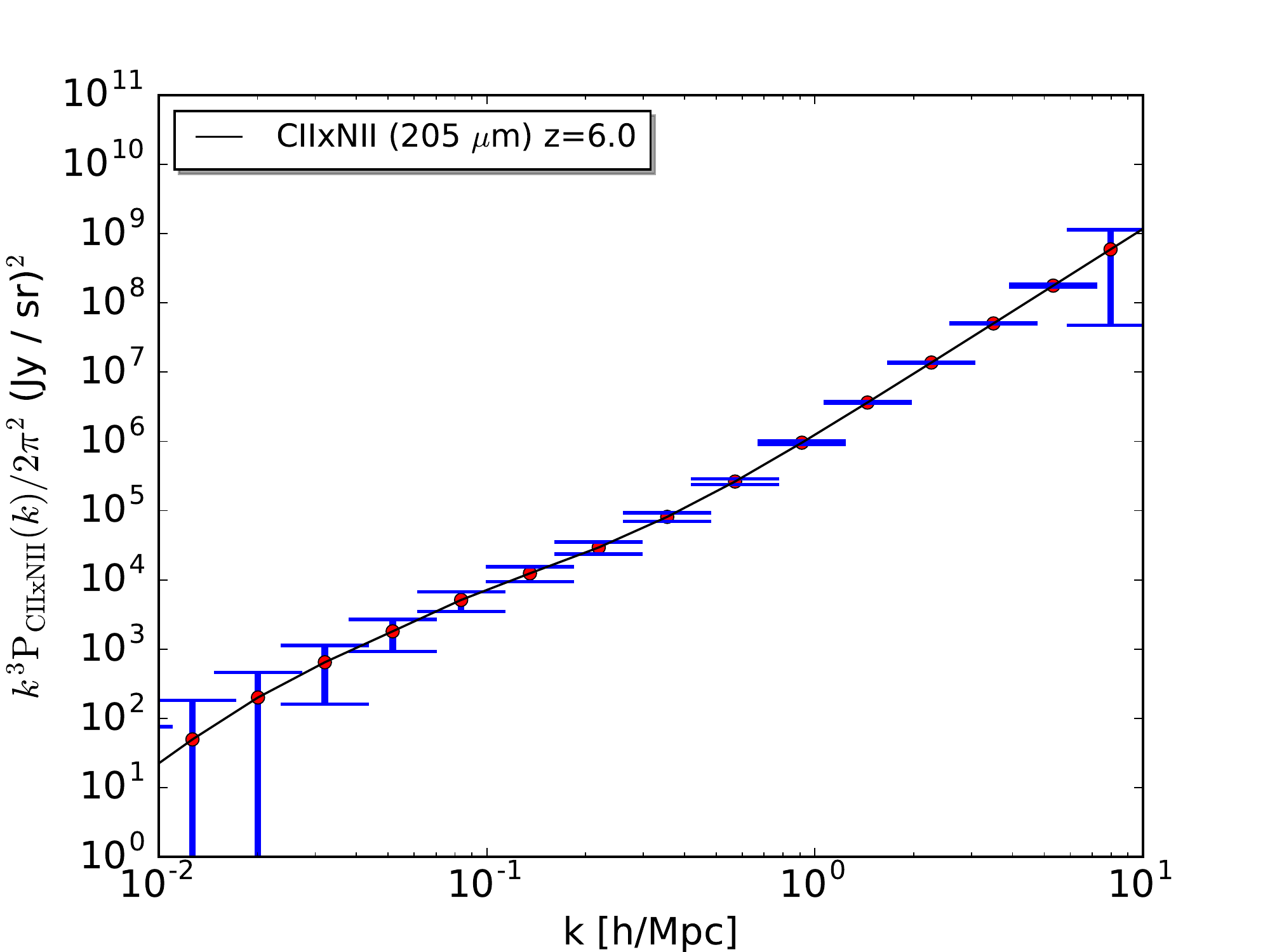}
\includegraphics[width=65mm]{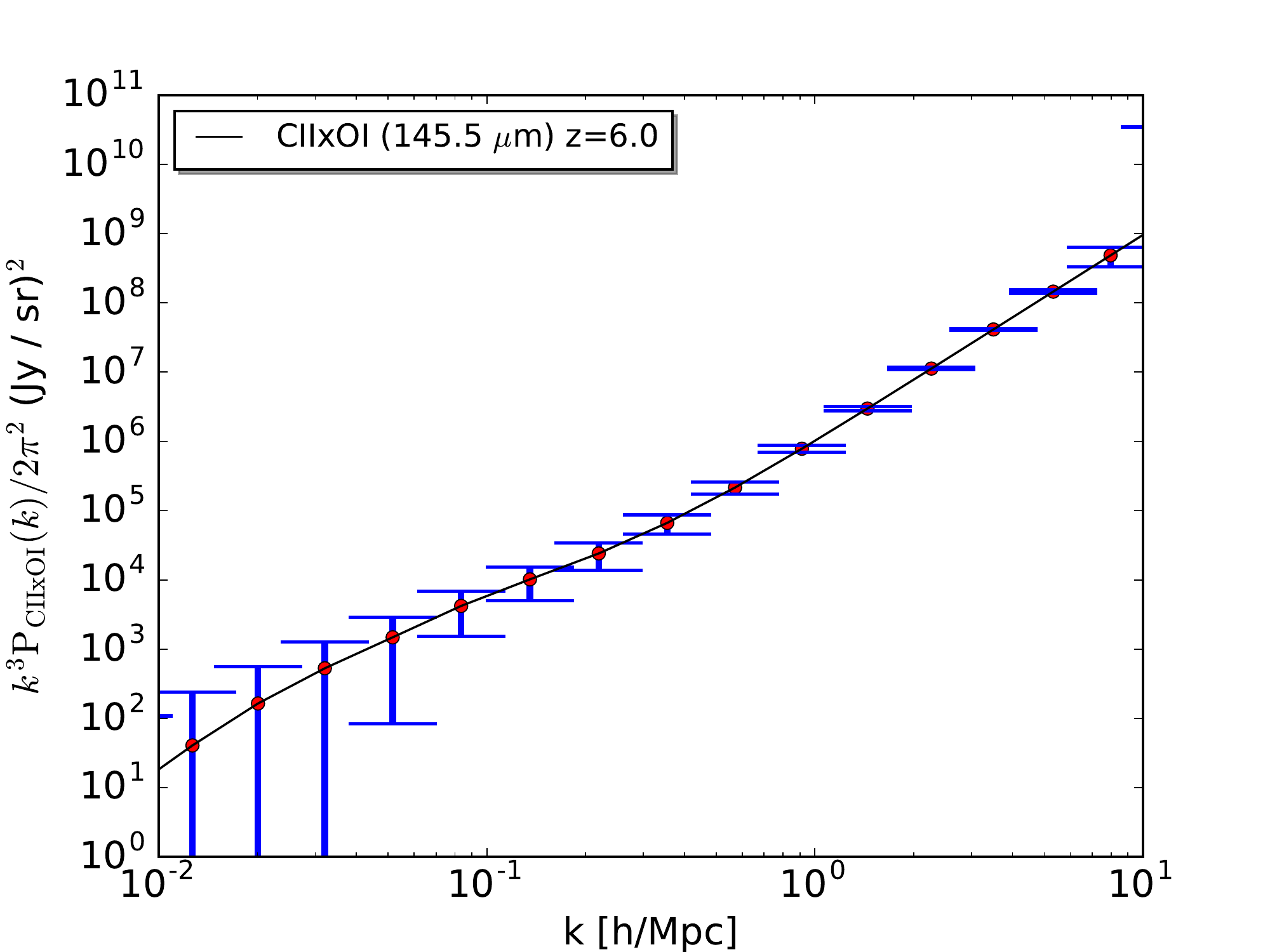}
\end{array} $
\end{center}
\caption{Predictions for both [CII] and total CO (3-2 to 7-6) auto-power spectra 
(left panel, black and green line respectively), and 
cross-spectra [CII]x[NII](205.2 $\mathrm{\mu}$m), and 
[CII]x[OI](145.5 $\mathrm{\mu}$m) at z=6 for a CII-stage II survey. All spectra are detected 
with high SNR.}
\label{fig:spectra_stage2_z6a}
\end{figure*}

\begin{table*}
\label{t4}
\centering
\begin{tabular}{|c|c|c|}
\cline{2-2}
 \hline
Instrument parameters & CONCERTO & CII-Stage II \\
     \hline
     Dish size (m) & 12 & 10 \\
     Survey Area (deg$^2$) & 2 & 100 \\
     Frequency range (GHz) & 200-360 & 200-300 \\
     Frequency resolution (GHz) & 1.5 & 0.4 \\
     Number of spectrometers & 1500 & 64  \\
     On-sky integration time (hr) & 1500 & 2000 \\
     NEFD on sky (mJy$\sqrt(sec)$& 155 & 5 \\
     \hline
     \end{tabular}
\caption{\label{tab:mcmc} Instrumental parameters for the two surveys, CONCERTO and CII-Stage II, considered.}
\end{table*}

The primary goal of the first survey considered, called CONCERTO, 
is to detect [CII] fluctuations in the redshift range 
$\mathrm{4.5<z<8.5}$. It is based on a spectrometer working in the
frequency range $\mathrm{200<\nu<360}$ GHz, with
spectral resolution $\mathrm{\delta_{\nu}\sim1.5}$ GHz. 
Such a frequency window imposes the use of a
so-called ``sub-millimetre'' telescope, with primary aperture size 
$\mathrm{D=12}$ m, and moderate angular resolution. 
The instrumental noise is thus computed for a total
observing time of $\mathrm{t_{survey} = 1500}$ hours, and a 
number of spectrometers $\mathrm{N_{sp}} = 1500$. 
The survey area considered here is two square degrees, 
and is optimized to ensure high SNR in the 
wavenumber range of $\mathrm{0.1<k<1}$ h / Mpc.\\
Accounting for realistic observational conditions and the 
total atmospheric transmission, the Noise Equivalent 
Flux density (NEFD), computed as the sensitivity 
per single pixel divided by the square root of the number of spectrometers, 
is equal to $\mathrm{NEFD=155}$ mJy\,sec$\mathrm{^{1/2}}$, 
for a spectral resolution of $\mathrm{\delta_{\nu}=1.5}$ GHz. 
The on-sky sensitivity $\mathrm{\sigma_{N}}$
can be expressed as:
\begin{eqnarray}
\sigma_{\mathrm{N}} = \frac{\mathrm{NEFD}}{\Delta\Omega_{\mathrm{beam}}}
\end{eqnarray}
 where
\begin{eqnarray}
\Delta\Omega_{\mathrm{beam}} &=& 2\pi \Big(\frac{\theta_{\mathrm{beam}}}{2.355}\Big)^2
\end{eqnarray}
is the beam area (in steradians), and the beam FWHM is given by:
\begin{eqnarray}
\theta_{\mathrm{beam}} &=& 1.22\lambda_{\mathrm{obs}} / D
\end{eqnarray}
where $\mathrm{\lambda_{\mathrm{obs}}}$ is the observed wavelength. 
Values for $\mathrm{\sigma_{N}}$ at $\mathrm{z=5}$, $\mathrm{z=6}$, and 
$\mathrm{z=7}$ are 15, 11, and 8.3 MJy/sr$\mathrm{\sqrt(sec)}$ respectively.
\\
The observing time per pixel is given by:
\begin{eqnarray}
t_{\mathrm{obs}} &=& t_{\mathrm{survey}}N_{\mathrm{sp}}\frac{\Delta\Omega_{\mathrm{pix}}}{\Delta\Omega_{\mathrm{survey}}},
\end{eqnarray}
where $\mathrm{\Delta\Omega_{survey}}$ is the total survey area covered.\\
Assuming a spherically averaged power spectrum measurement, and
a directionally independent on sky sensitivity $\sigma_{\mathrm{N}}$, 
the variance of the power spectrum is:
\begin{eqnarray}
\mathrm{var}[\bar{P}_{\alpha}(k)] = \frac{[P_{\alpha}(k)+
\bar{P}_{\alpha}^{\mathrm{N}}(k)]^2}{N_{\mathrm{m}}(k,z)},
\label{eqn:avg_spectrum}
\end{eqnarray}
where $\mathrm{N_m(k,z)}$ denotes the number of modes at each wavenumber:
\begin{eqnarray}
N_m(k,z) = 2\pi\,k^2\Delta\,k\frac{V_s}{(2\pi)^3};
\end{eqnarray}
the term $\mathrm{\Delta\,k}$ is the Fourier bin size, 
and V$_\mathrm{s}$(z) is the survey volume, expressed as:
\begin{eqnarray}
V_s(z) = \chi(z)^2\bar{y}\Delta\Omega_{\mathrm{survey}}B_{\nu}.
\end{eqnarray}
The averaged noise power spectrum in Eq.~\ref{eqn:avg_spectrum} is:
\begin{eqnarray}
\bar{P}_{\alpha}^{\mathrm{N}}(k) = V_{\mathrm{pix}} \frac{\sigma_N^2}{t_{\mathrm{obs}}};
\end{eqnarray}
where the volume surveyed by each pixel is: 
\begin{eqnarray}
V_{\mathrm{pix}}&=&\chi(z)^2\bar{y}_{\alpha}(z)\Omega_{\mathrm{beam}}\delta_\nu,
\end{eqnarray}
with
\begin{eqnarray}
\bar{y}_{\alpha}(z) &=& \lambda_{\alpha}(1+z)^2/H(z),
\end{eqnarray}
and $\mathrm{\lambda_{\alpha}}$ is the wavelength of the line $\alpha$ is the rest frame. \\
In Fig.~\ref{fig:spectra_concerto_z5} we plot measurements 
of the [CII] auto-power spectrum, 
together with [CII]x[OI] (145.5 $\mathrm{\mu}$m), and 
[CII]x[NII] (205.2 $\mathrm{\mu}$m) cross-power spectra 
at $\mathrm{z=5.0}$ for CONCERTO. 
For wavenumbers in the range $\mathrm{0.1<k<1}$ h / Mpc,
the [CII] auto-power spectrum will be detected with high significance ($\mathrm{SNR>50}$), while 
the [CII] cross-correlations with oxygen and nitrogen at these scales will not be very significant 
(SNR$\mathrm{\sim3}$ and SNR$\mathrm{\sim0.5}$ respectively). 
However, considering smaller scales (larger wavenumbers) the 
SNR increases significantly, and it will enable us to constrain the mean quantities  
$\mathrm{I_{[CII]}}$, $\mathrm{I_{[OI]}}$, and $\mathrm{I_{[NII]}}$.
Given the CONCERTO frequency coverage, at $\mathrm{z=6.0}$ it is possible to add the 
cross-correlation with [NII] (122 $\mathrm{\mu}$m). As shown in 
Fig.~\ref{fig:spectra_concerto_z6}, the cross-correlation of carbon with oxygen and 
nitrogen seems to be barely detectable at linear scales. 
However, in the non-linear regime, it might still 
be possible to measure these cross-correlations, and thus 
constrain the mean amplitude of these emission lines.
As already described in Sect.~\ref{sect:ism},
by looking at the cross-power spectra
[CII]x[NII] (121.9 $\mathrm{\mu)}$m, and [CII]x[NII]
(205.2 $\mathrm{\mu}$m), we would be able to measure the mean ratio
[NII] (205.2 $\mathrm{\mu}$m) / [NII] (121.9 $\mathrm{\mu}$m),
which is useful not only to constrain the electron density
of the low-ionized gas in HII regions,
but also to infer the mean emission of [CII] from PDRs,
and to constrain the global star formation rate.
The mean ratio between [OI] (145.5 $\mathrm{\mu}$m)
and CII is also a useful diagnostic of mean properties of
properties of PDRs, such as the hydrogen density and the
strength of the radiation field.\\
The second experimental setup, called CII-Stage II, has been introduced
in \cite{silva2015} as an appropriate baseline to ensure detection of [CII]
spectra in case of a pessimistic [CII] amplitude (see also \cite{lidz2016}).
It consists of a dish with diameter $\mathrm{D=10}$ m, with $\mathrm{16000}$
bolometers and $N_{\mathrm{sp}}=64$ beam spectrometers,
observing in the frequency range $\mathrm{200<\nu<300}$ GHz,
with a frequency resolution of 0.4 GHz. The total survey area is
100 deg$^2$ for a total observing time of t$_{\mathrm{survey}}=2000$ hours, and a
NEFD of 5 mJy\,sec$\mathrm{^{1/2}}$.\\
As it appears from Fig.~\ref{fig:spectra_stage2_z6a}, 
the cross-correlation of carbon with oxygen and nitrogen is now detectable with 
high SNR at $\mathrm{z=6}$. A space-based survey, being not limited by the atmosphere, would 
be able to perform measurements on a still wider frequency range, 
and thus perform measurements of high-redshift correlations 
with other interesting lines such as [OI] (63 $\mathrm{\mu}$m), 
[OIII] (88 $\mathrm{\mu}$m), [NIII] 57 $\mathrm{\mu}$m, and [CI]
(370 $\mathrm{\mu}$m and 609 $\mathrm{\mu}$m.
\begin{figure}[!b]
\begin{center}
\includegraphics[width=9.0cm]{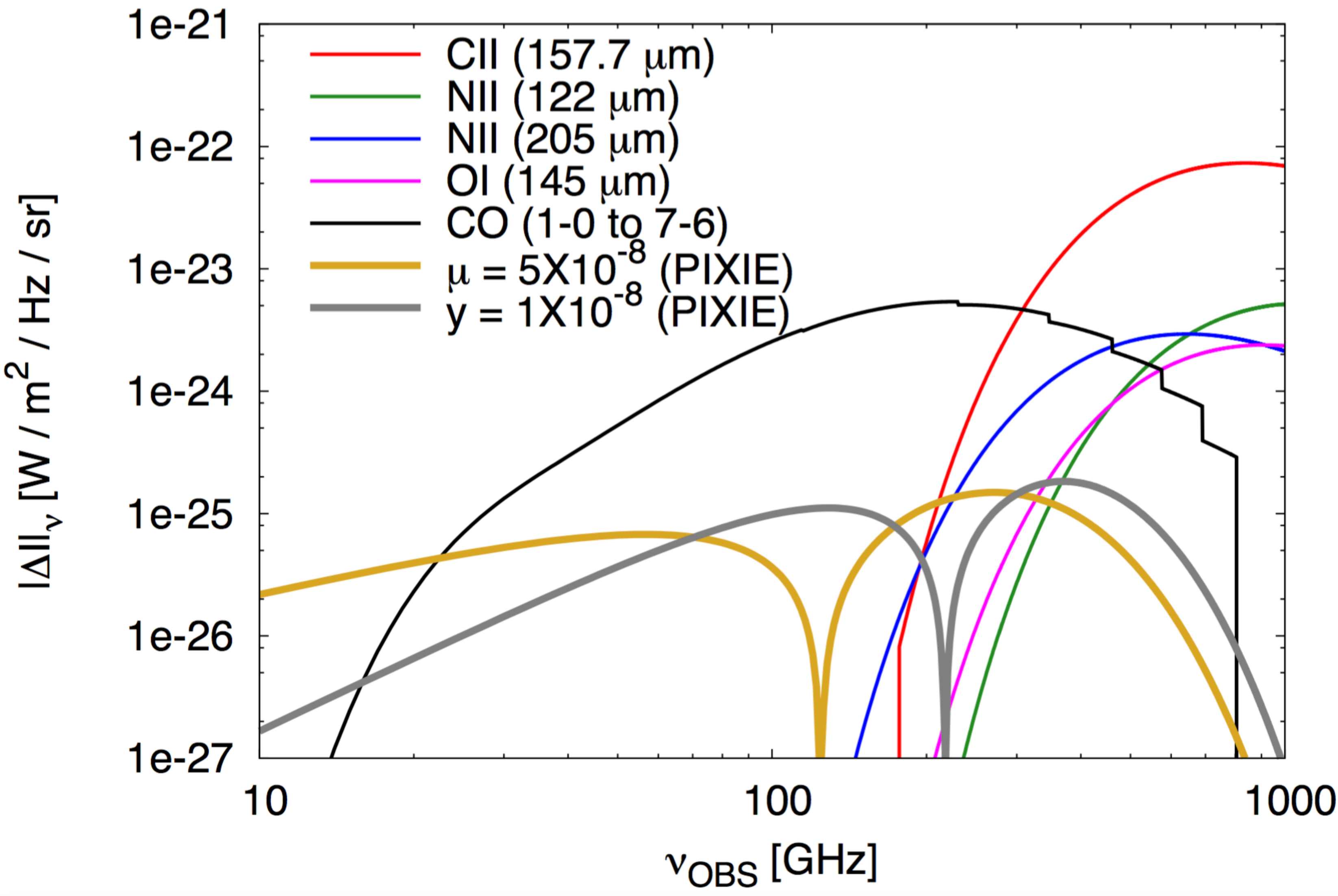}
\caption{Amplitudes of main emission lines that can be observed in the frequency range 
$\mathrm{\nu_{obs}=[10-1000]}$ GHz, 
together with the expected spectral distortions $\mu$ and $y$.}
\label{fig:em_lines}
\end{center}
\end{figure}

\section{Discussion}
\label{sect:discuss}
We have developed a consistent framework to compute predictions of 
3D power spectra of multiple FIR cooling lines of the ISM. Using measurements
of CIB power spectra, together with measurements of star formation
rate density from \cite{madau2014}, it is possible to constrain
the galaxy FIR luminosity at all redshift, which can be directly linked
to emission line amplitudes through scaling relation from
\cite{spinoglio2012}. Present and upcoming ground-based surveys 
aiming at measuring the power spectrum 
of the bright [CII] line, should be able to detect also the 
cross-correlation between the [CII] line and other lines 
produced in all phases of the ISM, such as 
[NII] (122 $\mathrm{\mu}$m and 205 $\mathrm{\mu}$m), 
and [OI] (145.5 $\mathrm{\mu}$m). Multiple measurements 
of cross-power spectra between [CII] and 
other emission lines will allow us to constrain 
the mean amplitude of 
each signal, and they will be key to 
gain insight into the mean properties of the ISM. 
Future surveys, such as PIXIE \citep{kogut2011,kogut2014}, 
working in a broad frequency range, 
will detect many more atomic and molecular lines emitted 
from moderate to high redshift with high SNR, 
allowing us to obtain multiple probes of 
all phases of the ISM. Moreover, the cross-correlation of the target line 
with galaxy number densities from future surveys such as, e.g., 
LSST \citep{lsst2009}, will be a powerful method to eliminate  
line foregrounds.\\
Line emissions from multiple atoms/molecules at 
multiple redshifts are also an important foreground for future 
surveys aiming at constraining CMB spectral 
distortions. In Fig.~\ref{fig:em_lines} we plot 
$\mathrm{\mu}$-type and y-type spectral distortions 
with $\mathrm{\mu = 5\cdot10^{-8}}$ and 
$\mathrm{y = 1\cdot10^{-8}}$, 
corresponding to the current PIXIE $\mathrm{5\sigma}$
sensitivity limits, together with the sum of the spectra 
from carbon monoxide emission lines 
(from $\mathrm{J=1\rightarrow0}$ to $\mathrm{J=7\rightarrow6}$), 
and the spectra from all emission lines considered in this work. 
The CO spectra have been computed using scaling 
relations from \cite{visbal2010} to link the CO line emission to 
the star formation rate, and the Kennicutt relation to express the 
star formation rate in terms of the galaxy 
infrared luminosity \citep{kennicutt1998}. The amplitude of the 
global signal from CO lines is similar to what found by \cite{mashian2016} 
using a radiative transfer modeling technique, 
even if the shape is slightly different.\\
We note that, even if foreground lines 
do not have a simple spectral dependence, unlike other foregrounds 
that can be modeled with power law such as synchrotron or thermal dust, 
their shape is still monotonic in frequency, and thus very 
different with respect to the CMB spectral
distortions. However, foreground subtraction will require a 
very good knowledge of the amplitude and shape of the total signal 
provided by the sum of these lines. The intensity 
mapping technique, by constraining the mean amplitude of the signal 
in multiple redshift bins, will help 
constraining the global contamination signal.\\
Finally, it is clear that an aggressive program to model 
the amplitude of all emission lines at all redshifts is necessary 
to have a detailed interpretation of upcoming measurements. 
Scaling relations are useful to work with, but 
they provide little information on the main physical mechanisms 
governing the line emission. Moreover, they are 
based on few observations performed at some given redshift, 
and their redshift evolution is not very well known. 
Different physical conditions can 
dominate the line emission at different epochs, 
strongly affecting the amplitude of the signal. As an example, at high 
redshift, the CMB strongly suppresses the [CII] emission 
from the cold neutral medium, leaving only the emission from PDRs 
\citep{vallini2015}. The redshift evolution of the galaxy infrared luminosity 
(which governs the evolution of the line emission in our model) 
is determined by the power law parameter $\delta$ (see Eq.~\ref{eqn:phiz}), 
which, as stated earlier, is quite 
uncertain, especially at high redshifts. 
On the other hand, semi-analytic models of galaxy formation and evolution 
often involve a large number of 
assumptions and free parameters, and such a complexity makes 
them difficult to use. A third approach, 
intermediate between the two, and based on present and upcoming measurements 
from, e.g. ALMA and SOFIA, should be developed to model the line intensity 
of all relevant emission lines, together with their redshift evolution. 
Such a model, possibly based on the physics of 
photodissociation regions, ionized medium, and molecular clouds, 
will offer an important guidance in interpreting upcoming and future 
intensity mapping observations, and thus constrain the mean 
properties of high-redshift galaxies.
 
%================================================================
{\acknowledgments
We thank the anonymous referee for many useful comments and suggestions. 
We thank Phil Bull, Tzu-Ching Chang, Abigail Crites, Roland de Putter and Paul Goldsmith 
for insightful discussions, and the organizers of the 
stimulating workshop ``Opportunities and Challenges in 
Intensity Mapping'' in Stanford. 
We acknowledge financial support 
from ``Programme National de Cosmologie and Galaxies'' 
(PNCG) of CNRS/INSU, France.
PS acknowledges hospitality from the 
Laboratoire d'Astrophysique de Marseille, 
where part of this work was completed. 
Part of the research described in this paper was carried out at the Jet Propulsion Laboratory, California Institute of Technology, under a contract with the National Aeronautics and Space Administration. Part of this work has been carried out thanks to the support of the OCEVU Labex (ANR-11-LABX-0060) and the A*MIDEX project (ANR-11-IDEX-0001-02) funded by the ``Investissements d'Avenir'' French government program managed by the ANR.}

\bibliography{astrobib}

%%%%%%%%%%%%%%%%%%%%%%%%%%%%%%%%%%%%%%%%%%%
% bibliography precedes appendices in A&A
%%%%%%%%%%%%%%%%%%%%%%%%%%%%%%%%%%%%%%%%%%%

%\appendix

\end{document}